\documentclass{article}

\usepackage[a4paper,
            left=2.3cm,right=2cm,top=2cm,bottom=2cm,footskip=.5cm]{geometry}
\usepackage{amsmath,amssymb}%
\usepackage{mathtools}%
\usepackage{authblk}
\usepackage{graphicx}%
\usepackage{hyperref}%
\usepackage{xcolor}%
	\DeclareMathOperator{\tr}{tr}%
	\DeclareMathOperator{\AiryAi}{Ai}%
	\DeclareMathOperator{\AiryBi}{Bi}%
	\DeclareMathOperator{\Riem}{\mathsf{Riem}}%
	\DeclareMathOperator{\Met}{\mathsf{Met}}%
	\DeclareMathOperator{\Diff}{\mathsf{Diff}}%
	
\newcommand*{\email}[1]{%
    \normalsize\href{mailto:#1}{#1}\par
    }
\begin{document}

\title{Quantum Phase Space Description of a Cosmological Minimal Massive Bigravity Model}%
\author[1]{Julio César Vera-Hernández}%

\affil[1]{
Departamento de Física, Escuela Superior de Física y Matemáticas\\
Instituto Politécnico Nacional, Unidad Profesional Adolfo López Mateos\\
Edificio 9, 07738, México D.F., México\\
\email{jcvera@ipn.mx} 
ORCID: 0000-0003-2167-4218
}

\maketitle%
\begin{abstract}
Bimetric gravity theories describes gravitational interactions in the presence of an extra spin-2 field. The Hassan-Rosen (HR) nonlinear massive minimal bigravity theory is a ghost-free bimetric theory formulated with respect a flat, dynamical reference metric. In this work the deformation quantization formalism is applied to a HR cosmological model in the minisuperspace. The quantization procedure is performed explicitly for quantum cosmology in the minisuperspace. The Friedmann-Lemaître-Robertson-Walker (FLRW) model with flat metrics is worked out and the computation of the Wigner functions for the Hartle-Hawking, Vilenkin and Linde wavefunctions are done numerically and, in the Hartle-Hawking case, also analytically. From the stability analysis in the quantum minisuper phase space it is found an interpretation of the mass of graviton as an emergent cosmological constant and as a measure of the deviation of classical behavior of the Wigner functions. Also, from the Hartle-Hawking case, an interesting relation between the curvature and the mass of graviton in a cusp catastrophe surface is discussed.\\

\textbf{Keyword}: Bigravity models, quantum cosmology, deformation quantization, Wigner functions
\end{abstract}

\section{\label{sec:Introduction}Introduction}

One of the most successful theories we have in physics is the Einstein's General Relativity Theory (GR) \cite{einstein_akad_1915}. We can observe that the majority of the phenomena related with gravity can be explained with the Einstein's Field Equations (EFE), from the classical Newtonian gravity force description to the light bend and gravitational waves observations: all of this can be simulated and reproduced in a proper limit with a very high precision.
Nevertheless, in standard cosmology there is a important problem related with the present acceleration of Universe. The usual explanation is consider the addition of a cosmological constant $\Lambda$ which must be adjusted to fit experimental data. An alternative is to look for modifications to GR in order to get weaker interaction at cosmological scales but that remains good to describe short distance behavior, and this is a important issue if we are interested in study the cosmology in the firsts instants of Big Bang. Many proposals for extend EFE has been developed over the years, where string theory is a well-know candidate for a quantum theory of gravity. GR is expressed as a non-linear self-interacting massless spin-2 particle field theory; from this point of view, a modification to gravity could be achieved through a massive spin-2 particle. Fierz and Pauli \cite{fierz_relativistic_1939} linearized a massive spin-2 field fluctuation, and Boulware and Deser showed that any nonlinear extension to Fierz-Pauli theory would exhibit a ghost inestability \cite{boulware_inconsistency_1972,boulware_can_1972}, but one of the most significant advance was the construction of a theory of ghost-free nonlinear massive gravity by de Rham, Gabadadze and Tolley (dRGT) \cite{de_rham_resummation_2011}. Later, Hassan and Rosen (HR) demonstrates that in a Hamiltonian constraint analysis, the Boulware-Deser ghost is abscent \cite{hassan_resolving_2012}.

dRGT theory develop the dynamics of a massive spin-2 field in a flat space in which two metric tensor components are playing roles, one as dynamical $g_{\mu\nu}$ and the other as a non-dynamical reference metric $f_{\mu\nu}$. As a result, open Friedmann-Lemaître-Robertson-Walker (FLRW) universes are allowed \cite{de_rham_stable_2014} but problems with strong coupling and ghostlike inestabilities arise \cite{yamashita_appearance_2014}. Attemping to avoid the problems with flat FLRW solutions, HR \cite{hassan_ghost-free_2012} extended the massive gravity theory beyond dRGT to a theory with two dynamical symmetric tensors $g_{\mu\nu}$ and $f_{\mu\nu}$, taked as foreground and background metrics, respectively, and having a symmetric and intertwined role. They called this a massive bigravity theory, and this shows a ghost-free description bimetric model containing nonlinear interactions of a massless and a massive spin-2 field in a dynamical background. In order to incorporate bimetric gravity models to cosmology, Rosen \cite{rosen_bimetric_1977} pointed the cosmological considerations of using bimetric theory with parametrized post-Newtonian formalism and giving examples of drastically different cosmologies emerging from bimetric theory. Later, Koenning \textit{et al.} \cite{koennig_stable_2014} discussed the stability of linear perturbations in an (nonlinear) ghost-free massive gravity, and exposing simple criteria for the cosmological stability of massive gravity for several classes of models, and Gümrükçüoğlu \textit{et al.} \cite{gumrukcuoglu_cosmology_2015} studied cosmological models in a bimetric theory with an effective composite coupling of matter, concluding that cosmological stability requires minimally coupled matter fields. In recent times, Lüben \textit{et al.} \cite{luben_bimetric_2020} claimed that bimetric cosmology is compatible with local tests of gravity, giving information about a viable cosmological expansion history.

The natural way to extend the classical bimetric cosmological theory to a quantum one is through canonical quantization. Nevertheless, if quantum mechanics is really the fundamental theory of whole nature, the study of quantum cosmology for flat or even open universes is necessary and deserves a formalism that performs the correct quantization of the model. Because the closed models have finite size and energy, they have usually taken as candidates for quantum cosmology models as they present finite action leading to a non-vanishing nucleation probability \cite{vilenkin_creation_1982}. In quantum cosmology framework, the whole universe is represented through a wavefunction satisfying the Wheeler-DeWitt equation \cite{dewitt_quantum_1967}. It is well-known as the low-energy approximation of string theory, however contains nontrivial information at leading order. In the beginning of the '80s of last century quantum cosmology born and one of the key ideas is that universe could be spontaneously created from nothing \cite{tryon_is_1973,vilenkin_creation_1982,rubakov_quantum_1984}. After nucleation, the universe can enter into a phase of inflationary expansion and continues its evolution to present time.

In order to obtain only a unique solution to Wheeler-DeWitt equation, it is necessary to impose boundary conditions, which is a problem in itself, because there is nothing external to universe. There are several proposals for the correct boundary conditions in quantum cosmology, among which the following are the most important: the no-boundary proposal of Hartle an Hawking \cite{hartle_wave_1983}, the tunneling proposal of Vilenkin \cite{vilenkin_quantum_1984,vilenkin_approaches_1994,vilenkin_cosmic_2001} and the proposal of Linde \cite{linde_quantum_1984}. Another important issue is how extract information of the Wheeler-DeWitt equation. In the general case, configuration space in quantum cosmology is an infinite-dimensional space called superspace and for its structure it is not suitable work with it. The next natural step is perform the reduction from superspace to minisuperspace, which is a more restricted framework where the set of spacelike geometries and matter fields are all but a finite set of components of spacelike metric and their conjugate momenta are put identically to zero \cite{pinto-neto_quantum_2013}. Even when this is not a rigorous scheme, the degree of space homogeneity of the primordial Universe suggest that this simplification can be physically feasible in quantum cosmology. There is the aim that minisuperspace maintains some essential features and properties of the quantum cosmology model \cite{kuchar_is_1989,halliwell_quantum_1991}.

As an important attempt to get information from Wheeler-DeWitt equation, at early 1990s quantum decoherence, the transition from quantum to classical physics was an active research area in quantum cosmology \cite{halliwell_how_2005}. In Ref. \cite{habib_wigner_1990} there is a development of quantum cosmology in the phase space and the use of Wigner function is shown to be a very useful approach to study decoherence. This is not a surprise because quantum mechanics in phase space is an appropiate formalism to describe quantum mechanical systems, especially those that admits descriptions of semiclassical properties, and the analysis of the classical limit is more clear in the Wigner function formalism. It has been argued that quantum decoherence is accomplished if the density matrix is averaged over phase space variables --the so-called coarse-grain density matrix approach. Nevertheless, the existence of classical correlation is another characteristic present in the classsical limit and requires the presence of clear, well-resolved extrema (maxima and minima), or peaks, of the Wigner function, and the use of coarse-grained density matrix produces a spreading of the distribution.

The quantum mechanics in phase space is only a part of a complete and mature type of quantization scheme: deformation quantization. In this formalism is assumed that the superspace (of 3-metrics) is flat and therefore a Fourier transform can be defined. This quantization emerges as a deformation of the usual product algebra of the smooth functions on the classical phase space and then as a deformation of the Poisson bracket algebra. The deformed product is called the $\star$-product which has been proved that exist in symplectic manifolds \cite{fedosov_simple_1994} and arbitrary Poisson manifolds \cite{kontsevich_operads_1999}. These results in principle allows to perform the quantization of arbitrary Poissonian or symplectic systems and to obtain the classical limit using the properties of $\star$-product. Due to the versatility of this scheme, the aim of this work is get a realization of the quantum cosmological structure for the model of minimal bigravity through the construction of the associated Wheeler-DeWitt-Moyal equation and to obtaining the Wigner function \cite{cordero_deformation_2011,cordero_phase_2019} for a zero-cuvature universe where the reference bigravity metric is negligible \cite{hassan_resolving_2012}, in order to extend the quantum cosmology of minimal bigravity models to phase spaces with non trivial topology. 

The structure of this paper is as follows. In Sec. \ref{sec:Canonical_Quantization_Minimal_Bigravity_Theory} the general ideas of massive bigravity theory are presented, aside its canonical quantization. In Sec. \ref{sec:Deformation_Quantization_Wigner_Function_Minimal_Bigravity_Model} the deformation quantization formalism for fields are presented, with the construction of Stratonovich-Weyl quantizer developed in \ref{sec:Stratonovich-Weyl_Quantizer}, the $\star$-product in Sec. \ref{sec:Moyal_Star_Product} and the Wigner functional in Sec. \ref{sec:Wigner_Functional}. In Sec. \ref{sec:Quantum_Cosmology_Minisuperspace} the quantum cosmology of a minimal bigravity model is worked in the minisuperspace: in Sec. \ref{sec:Canonical_Quantization_Minisuperspace} the Wheeler-DeWitt-Schrödinger is solved with the Hartle-Hawking, Vilenkin and Linde boundary conditions, in Sec. \ref{sec:Deformation_Quantization_Minisuperspace} the deformation quantization for the minisuperspace is introduced. Later, in Sec. \ref{sec:Exact_Solution_Hartle-Hawking_Case} the analytical Wigner function for the Hartle-Hawking is obtained toghether an analysis of the curvature and the mass of graviton, and in the Sec. \ref{sec:Numerical_Analysis} the numerical Wigner function for Hartle-Hawking, Vilenkin and Linde wavefunctions are calculated and the results interpreted. Finally, in Sec. \ref{sec:Conclusions} the conclusions and final remarks are provided.

\section{\label{sec:Canonical_Quantization_Minimal_Bigravity_Theory}Canonical Quantization of a Minimal Bigravity Theory}

The present section is an overview of the general procedure of a minimal massive gravity model. Only the principal results for the next work are presented, and detailed information can be found in Refs. \cite{de_rham_massive_2014,hassan_ghost-free_2012,hassan_resolving_2012,darabi_classical_2016}

\subsection{\label{sec:Hamiltonian_formalism}Hamiltonian formalism}

To set the Arnowitt--Deser-Misner (ADM) decomposition, is necessary consider a pseudo-Riemannian manifold $(\Omega,g)$ which is globally hyperbolic. Associated to this pseudo-Riemannian metric tensor $g$ there is another pseudo-Riemannian metric tensor $f$, the reference metric. In general dRGT theory, $f$ is necessary to ensure a ghost-free non-linear massive gravity theory. Thus, spacetime $\Omega$ admits the decomposition $\Omega = \mathbb R \times \Sigma$, where $\Sigma$ is a hypersurface on which the foliations $ds_g^2 = g_{\mu\nu}~dx^\mu dx^\nu = -(N^2 -N^i N_i)dt^2 + 2N_i dx^i dt + \gamma_{ij}dx^i dx^j$ and $ds_f^2 = f_{\mu\nu}~dx^\mu dx^\nu = -(M^2 -M^i M_i)dt^2 + 2M_i dx^i dt + \xi_{ij}dx^i dx^j$ are realized. Here $\gamma_{ij}$ and $\xi_{ij}$ are the extrinsic metrics on $\Sigma$, $N$ and $M$ the lapse functions, and $N^i$ and $M^i$ the shift vectors related to $g$ and $f$, respectively.

The space of all Riemannian $3$-metrics and scalar matter configurations $\phi$ defined on $\Sigma$ is called the \textit{superspace} $\Riem{\Sigma}=\{\gamma_{ij}(x),\phi(x):x\in\Sigma\}$. If $\Met(\Sigma)=\{\gamma_{ij}(x):x\in\Sigma\}$ denotes the infinite-dimensional space of Riemannian metrics on $\Sigma$, then the moduli space $\mathcal M$ is defined as the superspace modulo the group of diffeomorphisms $\Diff(\Sigma)$, $\mathcal M = \Riem(\Sigma)/\Diff(\Sigma)$ or for pure (bi)gravity, $\mathcal M = \Met(\Sigma)/\Diff(\Sigma)$.

The phase space for pure (bi)gravity (Wheeler's phase superspace) $\Gamma^*\cong T^*\Met(\Sigma)$, the cotangent bundle of $\Met(\Sigma)$, is given by the pairs $\Gamma^* = \{(\gamma_{ij},\pi^{ij}),(\xi_{ij},\varpi^{ij})\}$, where $\pi^{ij} = \frac{\partial L}{\partial \dot\gamma_{ij}}$, $\varpi^{ij} = \frac{\partial L}{\partial \dot\xi_{ij}}$, and $L$ is the HR bigravity massive Lagrangian. All this fields are supposed to be defined at $t = 0$, so $\gamma_{ij}(x,0) = \gamma_{ij}(x)$, $\xi_{ij}(x,0) = \xi_{ij}(x)$, $\pi^{ij}(x,0) = \pi^{ij}(x)$ and $\varpi^{ij}(x,0) = \varpi^{ij}(x)$.

The dynamics of minimal bigravity theory is encoded in the HR bigravity massive action which has the form
\begin{align}\label{eq:Hassan-Rosen_bigravity_action}
	\mathcal S &= M_g^2\int_\Omega d^4x~\sqrt{-\det g} ~R^{(g)} + M_f^2\int_\Omega d^4x~\sqrt{-\det f}~R^{(f)}\nonumber\\
						&+ 2m^2M_{\text{eff}}^2\int_\Omega d^4x~\sqrt{-\det g}\sum_{n=0}^4 \alpha_n L_n\left(\sqrt{g^{-1}f}\right) + \mathcal S_m,
\end{align}
where $g_{\mu\nu}$ and $f_{\mu\nu}$ are the two dynamical metric tensors components and $R^{(g)}$ and $R^{(f)}$ are their respective scalar curvatures in four dimensions. The matter action $\mathcal S_m = \int_\Omega d^4x \sqrt{\det g}\mathcal L_m(g,\phi)$ is the matter source containing a scalar field $\phi$, and for the case subject to a potential $\mathcal V(\phi)$, has the form
\begin{equation}
	\mathcal S_m = \int_\Omega d^4x~\sqrt{-\det g}\left( -\frac{1}{2} g^{\mu\nu}\partial_\mu \phi\partial_\nu \phi - \mathcal V(\phi)\right).
\end{equation}
The parameter $M_{\text{eff}}$ is an effective mass defined as the harmonic mean of squared field masses $M_g$ and $M_f$ associated to $g$ and $f$, respectively:
\begin{equation}\label{eq:effective_mass}
	\frac{1}{M_{\text{eff}}^2}=\frac{1}{M_g^2}+\frac{1}{M_f^2},
\end{equation}
and $m$ is the mass of graviton.

The meaning of the tensor $\sqrt{g^{-1}f}$ is such that
\begin{equation}\label{eq:bigravity_tensor}
	\left(\sqrt{g^{-1}f}\right)_\rho^\mu \left(\sqrt{g^{-1}f}\right)_\nu^\rho=g^{\mu\rho}f_{\rho\nu}=\chi^\mu_\nu,
\end{equation}
and the trace $\lfloor\chi\rfloor=\tr \chi^\mu_\nu$ serves to write the Lovelock-like Lagrangians \cite{hassan_ghost-free_2012,lovelock_einstein_1971,cruz_borninfeld_2013}
\begin{equation}
	\begin{gathered}\label{eq:Lovelock_Lagrangians}
		L_0(\chi)=1\\
		L_1(\chi)=\lfloor\chi\rfloor\\
		L_2(\chi)=\frac{1}{2}\left(\lfloor\chi\rfloor^2-\lfloor\chi^2\rfloor\right)\\
		L_3(\chi)=\frac{1}{6}\left(\lfloor\chi\rfloor^3-3\lfloor\chi\rfloor\lfloor\chi^2\rfloor+2\lfloor\chi^3\rfloor\right)\\
		L_4(\chi)=\frac{1}{24}\left(\lfloor\chi\rfloor^4-6\lfloor\chi\rfloor\lfloor\chi^2\rfloor+3\lfloor\chi^2\rfloor^2+8\lfloor\chi\rfloor\lfloor\chi^3\rfloor-6\lfloor\chi^4\rfloor\right)\\
		L_n(\chi)=0\qquad\text{for } n>4.
	\end{gathered}
\end{equation}

Using Eq. \eqref{eq:Lovelock_Lagrangians} and the HR minimal deformation of determinant structure expansion (\cite{hassan_non-linear_2011}, Eq.(3.3))
\begin{equation}
	3L_0\left(\left(\sqrt{g^{-1}f}\right)_\nu^\mu\right) - L_1\left(\left(\sqrt{g^{-1}f}\right)_\nu^\mu\right) + L_4\left(\left(\sqrt{g^{-1}f}\right)_\nu^\mu\right) = 3 - \tr\sqrt{g^{-1}f}+\det\sqrt{g^{-1}f}\label{eq:Hassan-Rosen_minimal_deformation},
\end{equation}
it is obtained the non-trivial minimal action \cite{darabi_classical_2016}
\begin{align}
	\mathcal S &= M_g^2\int_\Omega d^4x~\sqrt{-\det g}~R^{(g)} +M_f^2\int_\Omega d^4x~\sqrt{-\det f}~R^{(f)}\nonumber\\
											&+2m^2 M_{\text{eff}}^2\int_\Omega d^4x~\sqrt{-\det g}\left(3-\tr\sqrt{g^{-1}f}+\det\sqrt{g^{-1}f}\right)+\int_\Omega d^4x~\sqrt{-\det g}\mathcal L_m.
	\label{eq:non-trivial_minimal_action}
\end{align}

This action is highly non-linear in the lapse and shift variables. However, according with \cite{hassan_ghost-free_2012}, is possible get a linearized version with the introduction of three new variables $n^{i}$ which are functions of the lapse and shift variables and spatial metrics
\begin{equation}\label{eq:n_i_definition}
	N^{i} = M^{i} + (M\delta^{i}_{j} + N D^{i}_{j})n^{j}
\end{equation}

The matrix $D = D^{i}_{j}$ is given by
\begin{equation}\label{eq:D-matrix}
	D = \sqrt{\gamma^{-1}\xi Q}Q^{-1},
\end{equation}
where
\begin{equation}\label{eq:Q-matrix}
	Q^{i}_{j} = \alpha\delta^{i}_{j} + n^{i}n^{m}\xi_{mj}
\end{equation}
and
\begin{equation}\label{eq:alpha-scalar}
	\alpha = 1 - n^{k}\xi_{km} n^{m}.
\end{equation}

The matrix $D$ is independent of the lapses $N$ and $M$, and the shifts $N^{i}$ and $M^{i}$. With the replacement of $N^{i}$ in favor of $n^{i}$, the minimal action then becomes linear in the lapses $N$ and $M$, and the shift $M^{i}$:
\begin{align}\label{eq:D-shifted_action}
	\mathcal S &= \int_\Omega dt\,d^3x~\left\{ M_{g}^{2} \pi^{ij}\dot{\gamma}_{ij} + M_{f}^{2} \varpi^{ij}\dot{\xi}_{ij} + M^{i}\left(M_{g}^{2}R_{i}^{(g)} + M_{f}^{2}R_{i}^{(f)} + \pi_\phi \phi_{,i}\right)\right.\nonumber\\
	&+ M\left[M_{f}^{2}R^{0(f)} + M_{g}^{2} n^{i}R_{i}^{(g)} -2m^{2}M_{\text{eff}}^{2} (\sqrt{\det \gamma}\sqrt{\alpha} -\sqrt{\det \xi}) + \pi_\phi \phi_{,i}n^{i}\right]\nonumber\\
	&+ \left.N\left[M_g^2 R^{0(g)} + M_g^2 D^i_k n^k R_i^{(g)} - 2m^2 M_{\text{eff}}^2(\sqrt{\det \gamma}\sqrt{\alpha} D^k_k - 3\sqrt{\det\gamma}) \right.\right.\nonumber\\
	&+ \left.\left.\frac{\pi_\phi^2}{2\sqrt{\det \gamma}} + \frac{1}{2}\sqrt{\det\gamma}\gamma^{ij}\phi_{,i}\phi_{,j} + V(\phi) + D^{i}_{k}n^{k}\pi_\phi \phi_{,i}\right]\right\}
\end{align}
where
\begin{gather}\label{eq:R3g}
	R^{0(g)} = \sqrt{\det \gamma}\left[R^{3(g)} + \frac{1}{\det\gamma}\left(\frac{1}{2}\pi^2 - \pi^{ij}\pi_{ij}\right)\right],\\
	R^{i(g)} = 2\sqrt{\det\gamma}\nabla^{(g)}_j\left( \frac{\pi^{ij}}{\sqrt{\det\gamma}} \right),
\end{gather}
\begin{gather}\label{eq:R3f}
	R^{0(f)} = \sqrt{\det \xi}\left[R^{3(f)} + \frac{1}{\det\xi}\left(\frac{1}{2}\varpi^2 - \varpi^{ij}\varpi_{ij}\right)\right],\\
	R^{i(f)} = 2\sqrt{\det\xi}\nabla^{(f)}_j\left( \frac{\varpi^{ij}}{\sqrt{\det\xi}} \right),
\end{gather}
$\pi_\phi = \frac{\partial L}{\partial \dot\phi}$ and $,j$ denotes the derivative $\partial_j$. 
The canonical Hamiltonian can be written as
\begin{align}\label{eq:canonical_Hamiltonian}
	H_C &= \int_\Sigma d^3x~(\pi^{ij}\dot\gamma_{ij} + \varpi^{ij}\dot\xi_{ij} + \pi_\phi \dot\phi) - L\nonumber\\
			&= \int_\Sigma d^3x~(N\mathcal H_0^{(g)} + M\mathcal H_0^{(f)} + M^i\mathcal H_i^{(g)}),
\end{align}
and the corresponding equations of motion for $N$ and $M$ yields the $g,f$-Hamiltonian constraint and $M^{i}$ yield the momentum constraint:
\begin{align}\label{eq:general_Hamiltonian_constraint}
	\mathcal H_0 &= M_g^2 G_{ijkl}\pi^{ij}\pi^{kl} - M_g^2\sqrt{\det\gamma} R^{3(g)} - 2M_g^2 D^i_l n^l \gamma_{ij}\sqrt{\det\gamma} \nabla_k^{(g)}\left(\frac{\pi^{jk}}{\sqrt{\det\gamma}}\right) \nonumber\\
	&+ 2m^2M_{\text{eff}}^2 \sqrt{\det\gamma}(\sqrt{\alpha}D^k_k - 3) \nonumber\\
	&+ \frac{M}{N}\left\{M_f^2 F_{ijkl}\varpi^{ij}\varpi^{kl} - M_f^2 \sqrt{\det\xi} R^{3(f)} - 2M_f^2 n^{i}\xi_{ij}\sqrt{\det\xi} \nabla_k^{(f)}\left(\frac{\varpi^{jk}}{\sqrt{\det\xi}}\right) + \pi_\phi \phi_{,i}n^{i}\right\} \nonumber\\
	&+ \frac{\sqrt{\det\gamma}}{2}\left( \frac{\pi_\phi^2}{\det\gamma} + \gamma^{ij} \phi_{,i}\phi_{,j} + 2V(\phi) + 2D^{i}_{k} n^{k}\pi_\phi \phi_{,i}\right)
\end{align}
and
\begin{align}\label{eq:general_momentum_constraint}
	\mathcal H^{i} &= M_g^2\sqrt{\det\gamma} G^{0i} + M_f^2\sqrt{\det\xi} F^{0i} - \sqrt{\det\gamma} T^{0i}\nonumber\\
	&= - 2M_g^2 \sqrt{\det\gamma}\nabla_j^{(g)}\left( \frac{\pi^{ij}}{\sqrt{\det\gamma}} \right) - 2M_f^2 \sqrt{\det\xi}\nabla_j^{(f)}\left( \frac{\varpi^{ij}}{\sqrt{\det\xi}} \right) +\gamma^{ij}\pi_\phi \phi_{,j},
\end{align}
where $G_{ijkl} = \frac{1}{2}(\det\gamma)^{-1/2} (\gamma_{ik}\gamma_{jl} + \gamma_{il}\gamma_{jk} - \gamma_{ij}\gamma_{kl})$ and $F_{ijkl} = \frac{1}{2}(\det\xi)^{-1/2} (\xi_{ik}\xi_{jl} + \xi_{il}\xi_{jk} - \xi_{ij}\xi_{kl})$ are the $g$ and $f$ DeWitt metrics, respectively.

In this way, the Poisson brackets between canonical conjugate fields $\gamma_{ij}$ and $\pi^{ij}$ in one hand and $\xi_{ij}$ and $\varpi^{ij}$ by the other, are given by
\begin{gather}
	\{\gamma_{ij}(x), \pi^{kl}(y)\}_{\text{PB}} =\frac{1}{2}(\delta^k_i\delta^l_j + \delta^k_j\delta^l_i) \delta(x-y)\\
	\{\xi_{ij}(x), \varpi^{kl}(y)\}_{\text{PB}} =\frac{1}{2}(\delta^k_i\delta^l_j + \delta^k_j\delta^l_i) \delta(x-y),
\end{gather}
which are important structures in the process of quantization.

\subsection{\label{sec:Canonical_Quantization}Canonical quantization}

In order to apply the canonical quantization formalism in a minimal bigravity theory, is required that the promotion of canonical coordinates $\gamma_{ij}(x)$, $\xi_{ij}(x)$, $\phi(x)$, $\pi^{ij}$, $\varpi^{ij}$ and $\pi_\phi$ be done in the following form:
\begin{gather*}
	\hat{\gamma}_{ij}|\gamma_{ij},\xi_{ij},\phi\rangle = \gamma_{ij}|\gamma_{ij},\xi_{ij},\phi\rangle,\\
	\hat{\xi}_{ij}|\gamma_{ij},\xi_{ij},\phi\rangle = \xi_{ij}|\gamma_{ij},\xi_{ij},\phi\rangle,\\
	\hat{\phi}|\gamma_{ij},\xi_{ij},\phi\rangle = \phi(x)|\gamma_{ij},\xi_{ij},\phi\rangle,\\
	\hat{\pi}^{ij}|\gamma_{ij},\xi_{ij},\phi\rangle = -i\hslash\frac{\delta}{\delta \gamma_{ij}(x)}|\gamma_{ij},\xi_{ij},\phi\rangle,\\
	\hat{\varpi}^{ij}|\gamma_{ij},\xi_{ij},\phi\rangle = -i\hslash\frac{\delta}{\delta \xi_{ij}(x)}|\gamma_{ij},\xi_{ij},\phi\rangle,\\
	\hat{\pi}_\phi |\gamma_{ij},\xi_{ij},\phi\rangle = -i\hslash\frac{\delta}{\delta \phi(x)}|\gamma_{ij},\xi_{ij},\phi\rangle.
\end{gather*}
This operators satisfy the commutation relations
\begin{gather}\label{eq:general_commutation_relations}
	[\hat{\gamma}_{ij}(x),\hat{\pi}^{ij}(y)] = \frac{i\hslash}{2}(\delta^{k}_{i}\delta^{l}_{j} + \delta^{k}_{j}\delta^{l}_{i})\delta(x - y),\\
	[\hat{\xi}_{ij}(x),\hat{\varpi}^{ij}(y)] = \frac{i\hslash}{2}(\delta^{k}_{i}\delta^{l}_{j} + \delta^{k}_{j}\delta^{l}_{i})\delta(x - y).
\end{gather}
The constraints \eqref{eq:general_Hamiltonian_constraint} and \eqref{eq:general_momentum_constraint} have to be imposed at the quantum level in the form
\begin{equation}\label{eq:quantum_constraint}
	\hat{\mathcal H}_0|\Psi\rangle = 0,\qquad \hat{\mathcal H}_i|\Psi\rangle = 0
\end{equation}
In the coordinate representation, it follows that
\begin{align}
	&\left\{-M_g^2 G_{ijkl}\frac{\delta^2}{\delta\gamma_{ij}\delta\gamma_{kl}} - \frac{M}{N}M_f^2 F_{ijkl} \frac{\delta^2}{\delta\xi_{ij}\delta\xi_{kl}}\right.\nonumber\\
	&+\sqrt{\det\gamma}\left[-M_g^2 R^{3(g)} - 2M_g^2 D^i_l n^l \gamma_{ij} \nabla_k^{(g)}\left( \frac{\pi^{jk}}{\sqrt{\det\gamma}} \right) + 2m^2 M_{\text{eff}}^2 (\sqrt{\alpha}D^k_k -3)\right.\nonumber\\
	&+ \left.\left.\frac{M}{N}\left(- M_f^2 \sqrt{\det(\gamma^{-1}\xi)}R^{3(f)} - 2M_f^2 n^i \xi_{ij}\sqrt{\det(\gamma^{-1}\xi)} \nabla_k^{(f)}\left(\frac{\varpi^{jk}}{\sqrt{\det\xi}}\right)\right)\right] + \hat{T}^{00}\right\}\nonumber\\
	&\times\Psi[\gamma^{ij},\xi_{ij},\phi] = 0,
\end{align}
where $\langle\gamma_{ij},\xi_{ij},\phi|\Psi\rangle = \Psi[\gamma_{ij},\xi_{ij},\phi]$ and $\hat{T}^{00} = -\frac{1}{2}\frac{\delta^2}{\delta\phi^2} + \frac{1}{2}\gamma^{ij}\phi_{,i}\phi_{,j} + V(\phi) + \left(\frac{M}{N}n^i + D^i_k n^k\right) \pi_\phi \phi_{,i}$. This constraint is called the Wheeler-DeWitt equation and in general is not amenable to extract physical information of the system. it is a common practice reduce the number of degrees of freedom. In the following sections its considered a reduction of the model to one degree of freedom working in the so called minisuperspace.

\section{\label{sec:Deformation_Quantization_Wigner_Function_Minimal_Bigravity_Model}Deformation Quantization and Wigner Function of a Minimal Bigravity Model}

There are several methods to obtain the quantization of a classical system. As already made it in the previous section, the first choice is starting with the canonical quantization developed by Heisenberg, Schrödinger, Dirac and many others \cite{hall_quantum_2013}. By other way, with the path integral formulation established by Feynman \cite{feynman_space-time_1948}, another set of technical and conceptual advances has been acquired in formal quantization. Additional to this paradigms, another techniques has been constructed in the phase space, as the formalism of geometric quantization and deformation quantization, which emerges to deal with some problems of the previous methods \cite{bayen_deformation_1978}.

The Weyl-Wigner-Groenewold-Moyal (WWGM) formalism is an approach to the quantization of phase space \cite{weyl_quantenmechanik_1927,groenewold_principles_1946,moyal_quantum_1949}. Under this formalism, the Wigner function is the most important object; this function contains all the quantum information of the system. The key proposal of WWGM formalism is the connection between classical and quantum dynamics through the formal identification of usual canonical quantum observables with their classical counterparts.

With the aim of introduce the Weyl-Wigner-Groenewold-Moyal (WWGM) formalism for gravitational systems, first note that this is possible in a direct way since the minisuperspace in the minimal bigravity model is flat, and this ensures the existence of the Fourier transform, a key prerequisite. In this section a detailed description of WWGM formalism is left aside; the deformation quantization of gravity in ADM formalism and constrained systems, and in other cosmological models can be found in more detail in Refs. \cite{antonsen_deformation_1997,antonsen_deformation_1997-1,cordero_deformation_2011,cordero_phase_2019}.

\subsection{\label{sec:Stratonovich-Weyl_Quantizer}The Stratonovich-Weyl Quantizer}
Let $F[\gamma_{ij},\pi^{ij};\xi_{ij},\varpi^{ij};\phi,\pi_\phi]$ be a functional on the Wheeler phase superspace $\Gamma^*$ associated to hypersurface $\Sigma$ and let $\tilde F[\mu^{ij},\nu_{ij};\zeta^{ij},\theta_{ij};\kappa,\lambda]$ be its Fourier transform. By a pure analogy with the quantum structure, the Weyl quantization rule is defined as
\begin{align}\label{eq:Weyl_quantization_rule}
	\hat F&=\mathcal W(F[\gamma_{ij},\pi^{ij};\xi_{ij},\varpi^{ij};\phi,\pi_\phi])\nonumber\\
	:&=\int~\mathcal D\left(\frac{\nu_{ij}}{2\pi}\right)\mathcal D\left(\frac{\mu^{ij}}{2\pi}\right)\mathcal D\left(\frac{\theta_{ij}}{2\pi}\right)\mathcal D\left(\frac{\zeta^{ij}}{2\pi}\right)\mathcal D\left(\frac{\lambda}{2\pi}\right)\mathcal D\left(\frac{\kappa}{2\pi}\right)\nonumber\\
	&\times\tilde F[\mu^{ij},\nu_{ij};\zeta^{ij},\theta_{ij};\kappa,\lambda] \hat{\mathcal U}[\mu^{ij},\nu_{ij};\zeta^{ij},\theta_{ij};\kappa,\lambda],
\end{align}
where the functional measures are given by
\begin{equation}\label{eq:functional_measure}
	\mathcal D \gamma_{ij}=\prod_x d \gamma_{ij}(x)
\end{equation}
and so on, and the family of unitary operators 
\begin{equation}\label{eq:unitary_operators}
	\left\{\hat{\mathcal U}[\mu^{ij},\nu_{ij};\zeta^{ij},\theta_{ij};\kappa,\lambda] : (\mu^{ij},\nu_{ij}),(\zeta^{ij},\theta_{ij}),(\kappa,\lambda)\in \Gamma^*\right\}
\end{equation}
are expressed by
\begin{align}\label{eq:unitary_operator_realisation}
	\hat{\mathcal U}[\mu^{ij},\nu_{ij};\zeta^{ij},\theta_{ij};\kappa,\lambda]:&=\exp\left\{i\int dx~(\mu^{ij}(x)\hat\gamma_{ij}(x) + \nu_{ij}(x)\hat\pi^{ij}(x) + \zeta^{ij}(x)\hat{\xi}_{ij}\right.\nonumber\\
	&+ \left.\theta_{ij}(x)\hat\varpi^{ij}(x) + \kappa(x)\hat\phi(x) + \lambda(x)\hat{\pi}_\phi(x) )\right\},
\end{align}
with $\hat{\gamma}_{ij}, \hat{\pi}^{ij}, \hat{\xi}_{ij}, \hat{\varpi}^{ij}, \hat{\phi}$ and $\hat{\pi}_\phi$ field operators defined by
\begin{subequations}\label{eq:field_operators}
	\begin{align}
		\hat\gamma_{ij}(x)|\gamma_{ij},\xi_{ij},\phi\rangle&=\gamma_{ij}(x)|\gamma_{ij},\xi_{ij},\phi\rangle\label{eq:field_operators_a}\\
		\hat\pi^{ij}(x)|\pi^{ij},\varpi^{ij},\pi_\phi\rangle&=\pi_{ij}(x)|\pi^{ij},\varpi^{ij},\pi_\phi\rangle\label{eq:field_operators_b}\\
		\hat\xi_{ij}(x)|\gamma_{ij},\xi_{ij},\phi\rangle&=\xi_{ij}(x)|\gamma_{ij},\xi_{ij},\phi\rangle\label{eq:field_operators_e}\\
		\hat\varpi^{ij}(x)|\pi^{ij},\varpi^{ij},\pi_\phi\rangle&=\varpi^{ij}(x)|\pi^{ij},\varpi^{ij},\pi_\phi\rangle\label{eq:field_operators_f}\\
		\hat\phi(x)|\gamma_{ij},\xi_{ij},\phi\rangle&=\phi(x)|\gamma_{ij},\xi_{ij},\phi\rangle\label{eq:eq:field_operators_c}\\
		\hat{\pi}_\phi(x)|\pi^{ij},\varpi^{ij},\pi_\phi\rangle&=\pi_\phi(x)|\pi^{ij},\varpi^{ij},\pi_\phi\rangle\label{eq:field_operators_d}
	\end{align}
\end{subequations}

As it can be expected, this states form a complete basis while operators satisfies the usual canonical commutation relations. Using Campbell-Baker-Hausdorff theorem \cite{hall_lie_2015}, the completeness relations and standard formulas, Eq. \eqref{eq:unitary_operator_realisation} can be written in the form
\begin{align}\label{eq:bra-ket_unitary_representation}
	&\hat{\mathcal U}[\mu^{ij},\nu_{ij};\zeta^{ij},\theta_{ij};\kappa,\lambda] =\int\mathcal D \gamma_{ij} \mathcal D \xi_{ij}\mathcal D\phi \nonumber\\
	&\times\exp\left\{i\int dx~(\mu^{ij}(x) \gamma_{ij}(x) + \zeta^{ij}(x)\xi_{ij}(x) + \kappa(x)\phi(x)\right\}\nonumber\\
	&\times\left\lvert \gamma_{ij}-\frac{\hslash\nu_{ij}}{2}, \xi_{ij} - \frac{\hslash\theta_{ij}}{2},\phi-\frac{\hslash\lambda}{2}\right\rangle\left\langle \gamma_{ij} +\frac{\hslash\nu_{ij}}{2}, \xi_{ij} + \frac{\hslash\theta_{ij}}{2},\phi +\frac{\hslash\lambda}{2}\right\rvert,
\end{align}
where $\hslash$ denotes the deformation parameter which for non-null values resembles the deformation quantization. Following properties emerge from \eqref{eq:bra-ket_unitary_representation}:
\begin{equation}\label{eq:trace_unitary}
	\tr\{\hat{\mathcal U}[\mu^{ij},\nu_{ij};\zeta^{ij},\theta_{ij};\kappa,\lambda]\}=\delta[\nu_{ij}]\delta\left[\frac{\hslash\mu^{ij}}{2\pi}\right]\delta[\theta_{ij}]\delta\left[\frac{\hslash\zeta^{ij}}{2\pi}\right]\delta[\lambda]\delta\left[\frac{\hslash\kappa}{2\pi}\right]
\end{equation}
and
\begin{align}\label{eq:trace_unitary_product}
	&\tr\{\hat{\mathcal U}^\dagger[\mu^{ij},\nu_{ij};\zeta^{ij},\theta_{ij};\kappa,\lambda]\hat{\mathcal U}[\mu'^{ij},\nu'_{ij};\zeta'^{ij},\theta'_{ij};\kappa',\lambda']\} \nonumber\\
	&= \delta[\nu_{ij} - \nu'_{ij}] \delta[\theta_{ij} - \theta'_{ij}] \delta[\lambda - \lambda'] \nonumber\\
	&\times \delta\left[\frac{\hslash}{2\pi}(\mu^{ij} - \mu'^{ij})\right] \delta\left[\frac{\hslash}{2\pi}(\zeta^{ij} - \zeta'^{ij})\right] \delta\left[\frac{\hslash}{2\pi}(\kappa - \kappa')\right],
\end{align}
where $\tr \hat{\mathcal A}$ is the trace of operator $\hat{\mathcal A}$ in some convenient representation.

Using the previous results, the Weyl quantization rule can be written as
\begin{align}\label{eq:Stratonovich-Weyl_quantization_rule}
	\hat F &=\int \mathcal D\gamma_{ij} \mathcal D\left(\frac{\pi^{ij}}{2\pi\hslash}\right) \mathcal D\xi_{ij} \mathcal D\left(\frac{\varpi^{ij}}{2\pi\hslash}\right) \mathcal D\phi \mathcal D\left(\frac{\pi_\phi}{2\pi\hslash}\right) \nonumber\\
	&\times F[\gamma_{ij},\pi^{ij};\xi_{ij},\varpi^{ij};\phi,\pi_\phi]\hat\Omega[\gamma_{ij},\pi^{ij};\xi_{ij},\varpi^{ij};\phi,\pi_\phi],
\end{align}
where the operator
\begin{align}\label{eq:Stratonovich-Weyl_quantizer}
	&\hat\Omega[\gamma_{ij},\pi^{ij};\xi_{ij},\varpi^{ij};\phi,\pi_\phi]\nonumber\\
	&=\int \mathcal D\left(\frac{\hslash\nu_{ij}}{2\pi}\right)\mathcal D\mu^{ij} D\left(\frac{\hslash\theta_{ij}}{2\pi}\right)\mathcal D\zeta^{ij} \mathcal D\left(\frac{\hslash\lambda}{2\pi}\right)\mathcal D\kappa\mathcal \nonumber\\
	&\times\exp\left\{-i\int dx(\mu^{ij}(x)\gamma_{ij}(x) + \nu_{ij}(x)\pi^{ij}(x) + \zeta^{ij}(x)\xi_{ij}(x)\right.\nonumber\\
	&+\left.\vphantom{\int}\theta_{ij}(x)\varpi^{ij}(x) + \kappa(x)\phi(x)+\lambda(x)\pi_\phi(x))\right\}\times\hat{\mathcal U}[\mu^{ij},\nu_{ij};\zeta^{ij},\theta_{ij};\kappa,\lambda]
\end{align}
is the called Stratonovich-Weyl quantizer, which satisfies the relations
\begin{subequations}\label{eq:Stratonovich-Weyl_quantizer_properties}
	\begin{align}
		&\hat\Omega[\gamma_{ij},\pi^{ij};\xi_{ij},\varpi^{ij};\phi,\pi_\phi]^\dagger=\hat\Omega[\gamma_{ij},\pi^{ij};\xi_{ij},\varpi^{ij};\phi,\pi_\phi]\label{eq:Stratonovich-Weyl_quantizer_properties_a}\\
		&\tr(\hat\Omega[\gamma_{ij},\pi^{ij};\xi_{ij},\varpi^{ij};\phi,\pi_\phi])=1 \label{eq:Stratonovich-Weyl_quantizer_properties_b}\\
		&\tr\left(\hat\Omega[\gamma_{ij},\pi^{ij};\xi_{ij},\varpi^{ij};\phi,\pi_\phi] \hat\Omega[\gamma'_{ij},\pi'^{ij};\xi'_{ij},\varpi'^{ij};\phi',\pi'_\phi]\right)=\nonumber\\
		&\delta\left[\frac{\pi^{ij}-\pi'^{ij}}{2\pi\hslash}\right]\delta[\gamma_{ij}-\gamma'_{ij}] \delta\left[\frac{\varpi^{ij}-\varpi'^{ij}}{2\pi\hslash}\right]\delta[\xi_{ij}-\xi'_{ij}] \delta\left[\frac{\pi_\phi-\pi'_\phi}{2\pi\hslash}\right]\delta[\phi-\phi']\label{eq:Stratonovich-Weyl_quantizer_properties_c}.
	\end{align}
\end{subequations}

The Stratonovich-Weyl quantizer is useful to get the inverse map of $\mathcal W$:
\begin{equation}\label{eq:inverse_Weyl_rule}
	F[\gamma_{ij},\pi^{ij};\xi_{ij},\varpi^{ij};\phi,\pi_\phi]=\tr(\hat\Omega[\gamma_{ij},\pi^{ij};\xi_{ij},\varpi^{ij};\phi,\pi_\phi]\hat F),
\end{equation}
and using \eqref{eq:bra-ket_unitary_representation} on \eqref{eq:Stratonovich-Weyl_quantizer}, a most convenient form of Stratonovich-Weyl quantizer it is obtained:
\begin{align}\label{eq:Stratonovich-Weyl_quantizer_braket}
	&\hat\Omega[\gamma_{ij},\pi^{ij};\xi_{ij},\varpi^{ij};\phi,\pi_\phi] =\int\mathcal D\nu_{ij}~\mathcal D\theta_{ij}~\mathcal D\lambda \nonumber\\
	&\times\exp\left\{-\frac{i}{\hslash}\int dx( \nu_{ij}(x)\pi^{ij}(x) + \theta_{ij}(x)\varpi^{ij}(x) + \lambda(x)\pi_\phi(x))\right\}\nonumber\\
	&\times\left\lvert \gamma_{ij}-\frac{\nu_{ij}}{2}, \xi_{ij} - \frac{\theta_{ij}}{2}, \phi-\frac{\lambda}{2}\right\rangle\left\langle \gamma_{ij} + \frac{\nu_{ij}}{2}, \xi_{ij} + \frac{\theta_{ij}}{2}, \phi + \frac{\lambda}{2}\right\rvert.
\end{align}
\subsection{\label{sec:Moyal_Star_Product}Moyal Star Product}
The introduction of Moyal $\star$-product obey the need of a (noncommutative) product of usual functionals instead operators. Let $F=F[\gamma_{ij},\pi^{ij};\xi_{ij},\varpi^{ij};\phi,\pi_\phi]$ y $G=G[\gamma_{ij},\pi^{ij};\xi_{ij},\varpi^{ij};\phi,\pi_\phi]$ be two functionals defined on $\Gamma^*$ with associated field operators $\hat F$ and $\hat G$, respectively: 
\begin{align*}
F[\gamma_{ij},\pi^{ij};\xi_{ij},\varpi^{ij};\phi,\pi_\phi]&=\mathcal W^{-1}(\hat F)=\tr(\hat\Omega[\gamma_{ij},\pi^{ij};\xi_{ij},\varpi^{ij};\phi,\pi_\phi]\hat F)\\ G[\gamma_{ij},\pi^{ij};\xi_{ij},\varpi^{ij};\phi,\pi_\phi]&=\mathcal W^{-1}(\hat G)=\tr(\hat\Omega[\gamma_{ij},\pi^{ij};\xi_{ij},\varpi^{ij};\phi,\pi_\phi]\hat G).
\end{align*}

The functional related with operator product $\hat F\hat G$ is denoted by $(F\star G)[\gamma_{ij},\pi^{ij};\xi_{ij},\varpi^{ij};\phi,\pi_\phi]$ and have the realization
\begin{align}\label{eq:Moyal_star_product}
	&(F\star G)[\gamma_{ij},\pi^{ij};\xi_{ij},\varpi^{ij};\phi,\pi_\phi]\nonumber\\
	:&=\mathcal W^{-1}(\hat F\hat G)=\tr\{\hat\Omega[\gamma_{ij},\pi^{ij};\xi_{ij},\varpi^{ij};\phi,\pi_\phi]\hat F\hat G\}\nonumber\\
	&=F[\gamma_{ij},\pi^{ij};\xi_{ij},\varpi^{ij};\phi,\pi_\phi]\exp\left\{\frac{i\hslash}{2}\overleftrightarrow{\mathcal P}\right\}G[\gamma_{ij},\pi^{ij};\xi_{ij},\varpi^{ij};\phi,\pi_\phi],
\end{align}
where the bidiferential functional operator $\overleftrightarrow{\mathcal P}$ is defined by
\begin{align}\label{eq:bidiferential_operator}
	\overleftrightarrow{\mathcal P}:&=\int dx\left(\frac{\overleftarrow\delta}{\delta \gamma_{ij}(x)}\frac{\overrightarrow\delta}{\delta\pi^{ij}(x)}-\frac{\overleftarrow\delta}{\delta\pi^{ij}(x)}\frac{\overrightarrow\delta}{\delta \gamma_{ij}(x)}\right)\nonumber\\
	&=\int dx\left(\frac{\overleftarrow\delta}{\delta \xi_{ij}(x)}\frac{\overrightarrow\delta}{\delta\varpi^{ij}(x)}-\frac{\overleftarrow\delta}{\delta\varpi^{ij}(x)}\frac{\overrightarrow\delta}{\delta \xi_{ij}(x)}\right)\nonumber\\
	&+\int dx\left(\frac{\overleftarrow\delta}{\delta\phi(x)}\frac{\overrightarrow\delta}{\delta\pi_\phi(x)}-\frac{\overleftarrow\delta}{\delta\pi_\phi(x)}\frac{\overrightarrow\delta}{\delta\phi(x)}\right).
\end{align}
The Eq. \eqref{eq:Moyal_star_product} is know as the Moyal $\star$-product for fields.

\subsection{\label{sec:Wigner_Functional}The Wigner Functional}
The fundamental concept in deformation quantization formalism is the Wigner function, so the Wigner functional takes the same role in a straightforward generalization for infinite degrees of freedom. If $\hat\rho$ is the density operator associated to a quantum state, then the Wigner functional $\rho[\gamma_{ij},\pi^{ij};\xi_{ij},\varpi^{ij};\phi,\pi_\phi]$ corresponding to $\hat\rho$ reads as
\begin{align}\label{eq:Wigner_functional}
	&\rho[\gamma_{ij},\pi^{ij};\xi_{ij},\varpi^{ij};\phi,\pi_\phi]=\tr\{\hat\Omega[\gamma_{ij},\pi^{ij};\xi_{ij},\varpi^{ij};\phi,\pi_\phi]\hat\rho\}\nonumber\\
	&=\int\mathcal D\left(\frac{\nu_{ij}}{2\pi\hslash}\right)~\mathcal D\left(\frac{\theta_{ij}}{2\pi\hslash}\right)~\mathcal D\left(\frac{\lambda}{2\pi\hslash}\right) \nonumber\\
	&\times \exp\left\{-\frac{i}{\hslash}\int dx( \nu_{ij}(x)\pi^{ij}(x) + \theta_{ij}(x)\varpi^{ij}(x) + \lambda(x)\pi_\phi(x))\right\}\nonumber\\
	& \times\left\langle \gamma_{ij} + \frac{\nu_{ij}}{2}, \xi_{ij} + \frac{\theta_{ij}}{2}, \phi + \frac{\lambda}{2}\right\rvert \hat{\rho} \left\lvert \gamma_{ij}-\frac{\nu_{ij}}{2}, \xi_{ij} - \frac{\theta_{ij}}{2}, \phi-\frac{\lambda}{2}\right\rangle.
\end{align}

In case of a pure state of the system $\hat\rho=\lvert\Psi\rangle\langle\Psi\rvert$, Eq. \eqref{eq:Wigner_functional} implies
\begin{align}\label{eq:Wigner_functional_wavefunctions}
	&\rho[\gamma_{ij},\pi^{ij};\xi_{ij},\varpi^{ij};\phi,\pi_\phi]=\tr\{\hat\Omega[\gamma_{ij},\pi^{ij};\xi_{ij},\varpi^{ij};\phi,\pi_\phi]\hat\rho\}\nonumber\\
	&=\int\mathcal D\left(\frac{\nu_{ij}}{2\pi\hslash}\right)~\mathcal D\left(\frac{\theta_{ij}}{2\pi\hslash}\right)~\mathcal D\left(\frac{\lambda}{2\pi\hslash}\right)\nonumber\\
	&\times \exp\left\{-\frac{i}{\hslash}\int dx( \nu_{ij}(x)\pi^{ij}(x) + \theta_{ij}(x)\varpi^{ij}(x) + \lambda(x)\pi_\phi(x))\right\}\nonumber\\
	& \times \Psi^*\left[\gamma_{ij}-\frac{\nu_{ij}}{2}, \xi_{ij} - \frac{\theta_{ij}}{2}, \phi-\frac{\lambda}{2}\right] \Psi\left[\gamma_{ij} + \frac{\nu_{ij}}{2}, \xi_{ij} + \frac{\theta_{ij}}{2}, \phi + \frac{\lambda}{2}\right].
\end{align}
where $\langle \gamma_{ij},\xi_{ij},\phi\rvert\Psi\rangle=\Psi[\gamma_{ij}, \xi_{ij},\phi]$ is the wavefuntion of universe.

The expected value for a general operator $\tilde F$ can be expressed through $\hat\rho$ as
\begin{align}\label{eq:expected_value}
\langle\hat F\rangle&=\frac{\tr(\hat\rho\hat F)}{\tr\hat\rho}\nonumber\\
&=\left\{\int \mathcal D\gamma_{ij} \mathcal D\left(\frac{\pi^{ij}}{2\pi\hslash}\right) \mathcal D\xi_{ij} \mathcal D\left(\frac{\varpi^{ij}}{2\pi\hslash}\right) \mathcal D\phi \mathcal D\left(\frac{\pi_\phi}{2\pi\hslash}\right)\right.\nonumber\\
&\times\left.\vphantom{\int}\rho[\gamma_{ij},\pi^{ij};\xi_{ij},\varpi^{ij};\phi,\pi_\phi]\tr(\hat\Omega[\gamma_{ij},\pi^{ij};\xi_{ij},\varpi^{ij};\phi,\pi_\phi]\hat F)\right\}\nonumber\\
&\left/\vphantom{\int\mathcal D}\middle\{\int\mathcal D\pi^{ij}\mathcal D\gamma_{ij}\mathcal D\pi_\phi\mathcal D\phi~\rho[\gamma_{ij},\pi^{ij};\xi_{ij},\varpi^{ij};\phi,\pi_\phi]\right\}.
\end{align}

Follow the prescription to construct the equations of evolution of the quantum system, it is possible to write the equivalent of Hamiltonian constraint in terms of Wigner functional and Moyal $\star$-product as
\begin{equation}\label{eq:Wheeler-DeWitt-Moyal}
	\mathcal H_0 \star \rho[\gamma_{ij},\pi^{ij};\xi_{ij},\varpi^{ij};\phi,\pi_\phi]=0
\end{equation}
Eq. \eqref{eq:Wheeler-DeWitt-Moyal} is called the Wheeler-DeWitt-Moyal equation, and is the deformation quantization version of Wheeler-DeWitt-Schrödinger equation. The first point take into account is that the dynamics of the system is entirely determined by this equation and their explicit form depends of the particularity of the problem. 

\section{\label{sec:Quantum_Cosmology_Minisuperspace}Quantum Cosmology of a Minimal Bigravity Model in the Minisuperspace}

Although the general construction exposed in Secs. \ref{sec:Canonical_Quantization_Minimal_Bigravity_Theory} and \ref{sec:Deformation_Quantization_Wigner_Function_Minimal_Bigravity_Model} in terms of functional integrals defined in the whole superspace of three-metrics is necessary in the more general case, the restriction to work in minisuperspace is natural due to its simplicity and because has been widely studied in the literature by other methods. The principal goal is to obtain the Wigner function for a minimal bigravity model and to motivate a further study using deformation quantization. This first step is necessary to gain some experience to eventually extend the deformation quantization to curved phase spaces. It is known that deformation quantization formalism admits a natural extension to these scenarios and allows suitable generalizations.

\subsection{\label{sec:Minimal_Bigravity_Cosmology_Minisuperspace}Minimal Bigravity Cosmology in the Minisuperspace}

For the development of a standard cosmological theory, it is need to obtain the point-like Hamiltonian of this theory assuming two Friedmann-Lemaître-Robertson-Walker-like (FLRW) dynamical line elements such that they are $SO(4)$ invariant metrics in a $\Omega = \mathbb R\times S^3$ topology in a homogeneus and isotropic spacetime:

\begin{equation}\label{eq:g-FLRW}
	ds_g^2 = l_p^2 \left[-N^2(t)~dt^2 + a^2(t) \left(\frac{dr^2}{1-\kappa r^2} + r^2~d\Omega^2\right)\right]
\end{equation}
and
\begin{equation}\label{eq:f-FLRW}
	ds_f^2 = l_p^2 \left[-M^2(t)~dt^2 + b^2(t) \left(\frac{dr^2}{1-\kappa r^2} + r^2~d\Omega^2\right)\right].
\end{equation}
where $N(t)$ and $M(t)$ are the lapse functions, $a(t)$ and $b(t)$ are the scale factors of metrics $g_{\mu\nu}$ and $f_{\mu\nu}$, respectively, $\kappa$ is the space curvature, assumed to be the same for both metrics, and $l_p = 2/3 L_p$, where $L_p$ denotes the Planck length.

So, using the expressions of $R^{3(g)} = 6\kappa/a^2$, $R^{3(f)} = 6\kappa/b^2$ and the metrics above on the field equations, and using the assumption $M_g^2=M_f^2=M_{\text{eff}}^2/2$, it is obtained a point-like gravitational Hamiltonian in the minisuperspace $\{N,a,M,b\}$:
\begin{align}\label{eq:Hamiltonian_constraint}
	\mathcal H &= N\mathcal H_0\nonumber\\
	&= N \left\{-\frac{p_a^2}{4 a} - \kappa a - 2\mu^2 a^3 \left( 1-\frac{b}{a} \right)\right.\nonumber\\
	&+\left. \frac{M}{N}\left[-\frac{p_b^2}{4 b} - \kappa b - \frac{2\mu^2}{3}a^3\left(\frac{b^3}{a^3} - 1\right)\right]\right\} + \mathcal H_m,
\end{align}
where $\mu:=m l_p$ and $\mathcal H_m$ is the matter Hamiltonian. As mentioned in Sec. \ref{sec:Hamiltonian_formalism}, in ADM formalism the Hamiltonian constraint can be written as $\mathcal H_0=0$. In canonical quantization, this constraint in the form of a operator $\hat{\mathcal H}$ is the Wheeler-DeWitt equation, where the wavefunction of the universe $\Psi$ belongs to $\ker \hat{\mathcal H}$.

On the other hand, it can be noted that the matter term in the action is independent of the modifications of the bigravity model by mass term. So, the matter part of the Hamiltonian can simply added to the geometric one to obtain the total Hamiltonian expression. In order to model the matter part Hamiltonian, the equation of state for a perfect fluid matter it is used:
\begin{equation}\label{eq:perfect_fluid}
	P = \omega \rho.
\end{equation}
In the Schutz representation \cite{schutz_perfect_1970,vakili_quadratic_2010} for a perfect fluid, the matter part Hamiltonian can be expressed as
\begin{equation}\label{eq:matter_Hamitonian}
	\mathcal H_m=\frac{Np_T}{a^{3\omega}},
\end{equation}
where the conjugate momentum $p_T$ associated to dynamical variable $T$ (the collective thermodynamic parameters of perfect fluid) shows a convenient interpretation as a time-related magnitude. So, including this as dynamical parameter, the Hamiltonian constraint is expressed as
\begin{equation}\label{eq:massive_Hamiltonian_constraint}
	\mathcal H_0 = -\frac{p_a^2}{4 a} - \kappa a - 2\mu^2a^3 \left(1-\frac{b}{a}\right) +\frac{p_T}{a^{3\omega}} + \frac{M}{N}\left[-\frac{p_b^2}{4 b} - \kappa b - \frac{2\mu^2 a^3}{3}\left( \frac{b^3}{a^3} - 1\right)\right].
\end{equation}
As it is seen from construction, only the matter is coupled with the metric $g_{\mu\nu}$.

Taking the Hamilton canonical classical equations
\begin{equation}\label{eq:Hamilton-a}
	\dot a = \{a,\mathcal H\}_{\text{PB}} = -\frac{N p_a}{2a}
\end{equation}
and
\begin{equation}\label{eq:Hamilton-b}
	\dot b = \{b,\mathcal H\}_{\text{PB}} = -\frac{M p_b}{2b}
\end{equation}
it is possible to calculate the equations of motion corresponding to the minisuperspace variables $N$ and $M$:
\begin{equation}\label{eq:g-pre-Hubble}
	\frac{\dot a^2}{N^2a^2} + \frac{\kappa}{a^2} + 2\mu^2\left(1 - \frac{b}{a}\right)=0,
\end{equation}
\begin{equation}\label{eq:f-pre-Hubble}
	\frac{\dot b^2}{M^2b^2} + \frac{\kappa}{b^2} + \frac{2\mu^2}{3}\left(1 - \frac{a^3}{b^3}\right)=0,
\end{equation} 
 or defining the Hubble-like parameters $H=\dot a/Na$ and $K=\dot b/b$,
\begin{equation}\label{eq:g-Hubble}
	H^2+\frac{\kappa}{a^2}+2\mu^2\left(1-\frac{b}{a}\right)=0,
\end{equation}
\begin{equation}\label{eq:f-Hubble}
	K^2 + \frac{\kappa M^2}{b^2} + \frac{2}{3}\mu^2 M^2 \left(1 - \frac{a^3}{b^3}\right)=0.
\end{equation}
There is an interesting discussion about Eqs. \eqref{eq:g-Hubble} and \eqref{eq:f-Hubble}. A comparision with the standard Friedmann equation reveals that the reduced mass of graviton $\mu^2$ plays the role of a comological constant $\Lambda$, provided $\frac{b}{a}$ takes a constant value, which can be positive (negative) for $b > a$ and negative (positive) if $b < a$. This emerging parameter can be interpret the dynamical competition between the two scale factors as a phase transition from deceleration to acceleration era. So, to taking in account this point, is natural to make the assumption $\sigma:=b/a=\text{const}$ and, with the use of the momentum constraint \eqref{eq:general_momentum_constraint}, write
\begin{equation}\label{eq:second_Bianchi_constraint_quotient}
	\sigma=\frac{b}{a}=\frac{\dot b}{\dot a}=\frac{M}{N},
\end{equation}
and then, $p_b = \sigma p_a$. With all this, Eq. \eqref{eq:massive_Hamiltonian_constraint} then can be written as
\begin{equation}\label{eq:sigma-Hamiltonian_constraint}
	\mathcal H_0 = - (1 + \sigma^2)\frac{p^2}{4a} - (1 + \sigma^2)\kappa a -2\mu^2 a^3\left( 1 - \frac{4\sigma}{3} + \frac{\sigma^4}{3}\right) + \frac{p_T}{a^{3\omega}},
\end{equation}
where $p := p_a$ for simplicity.

\subsection{\label{sec:Canonical_Quantization_Minisuperspace}Canonical Quantization in the Minisuperspace}

In the canonical quantization formalism, the Wheeler-DeWitt equation $\hat{\mathcal H}_0\Psi=0$ is the operator version of the Hamiltonian constraint $\mathcal H_0=0$; the solution of this equation is the wavefunction of universe $\Psi$, which encodes all the information of the evolution of universe. So, after the replacements $p \mapsto \hat{p}$ and $p_T\mapsto\hat{p}_T$ over Eq. \eqref{eq:sigma-Hamiltonian_constraint}, the Wheeler-DeWitt reads as
\begin{align}\label{eq:Wheeler-DeWitt}
	\hat{\mathcal H}_0\Psi(a,T) &= \left\{- (1 + \sigma^2)\frac{\hat{p}^2}{4a} - (1 + \sigma^2)\kappa a\right.\nonumber\\
	&\left. -2\mu^2 a^3\left( 1 - \frac{4\sigma}{3} + \frac{\sigma^4}{3}\right) + \frac{\hat{p}_T}{a^{3\omega}}\right\}\Psi(a,T)=0.
\end{align}

Eq. \eqref{eq:Wheeler-DeWitt} admits the separation
\begin{equation}\label{eq:T-separation}
	\Psi(a,T) = e^{iET}\psi(a),
\end{equation}
with $E$ a constant. With this, Wheeler-DeWitt equation takes the form
\begin{equation}\label{eq:Wheeler-DeWitt-Schrodinger}
	\left[-(1+\sigma^2)\frac{\hat{p}_a^2}{4a} - (1+\sigma^2)\kappa a +\frac{E}{a^{3\omega}} - 2\mu^2a^3\left(1-\frac{4}{3}\sigma+\frac{\sigma^4}{3}\right)\right]\psi(a)=0.
\end{equation}

As it can see, the coupling between the matter source and the gravitation $g_{\mu\nu}$ states the identification of two metrics $g_{\mu\nu}$ and $f_{\mu\nu}$ and this reduces from two-variable system to one-variable system in favor just the scale factor $a$, getting a formal reduction $\Psi(a,b)\equiv \Psi(a,\sigma)\equiv \psi(a)$. To avoid further confusion, Eq. \eqref{eq:Wheeler-DeWitt-Schrodinger} is referred as Wheeler-DeWitt-Schrödinger (WDWS) equation.

\subsubsection{\label{sec:Exact_Solution}Exact Solution}

As mentioned by Darabi and Mousavi \cite{darabi_classical_2016}, in this representation the conjugate momentum associated with $T$ appears linearly in the Hamiltonian of the model, so without loss of generality $T$ can take the role of time-variable in the Wheeler-DeWitt-Schrödinger equation \eqref{eq:Wheeler-DeWitt-Schrodinger}. The emergency of a Schrödinger-like equation in this model give us useful information about $T$-parameter system evolution. On the other hand, through $T$ the factor order operator for the operators $\hat a$ and $\hat p_a$ can be considered by means of
\begin{equation}\label{eq:order_operator}
	\hat p_a^2= - a^{-\lambda}\dfrac{\partial}{\partial a}\left( a^{\lambda} \dfrac{\partial}{\partial a}\right).
\end{equation}
Using Eq. \eqref{eq:order_operator}, the next equation it is obtained:
\begin{equation}\label{eq:T-independent_Wheeler-DeWitt-Schrodinger}
	\left[\frac{1}{4a}(1+\sigma^2)\left( \frac{d^2}{da^2} + \frac{\lambda}{a}\frac{d}{da} \right) - (1+\sigma^2)\kappa a + \frac{E}{a^{3\omega}} - 2\mu^2a^3\left(1-\frac{4}{3}\sigma + \frac{\sigma^4}{3}\right)\right]\psi(a) = 0
\end{equation}
Taking $\omega = -1$ and
\begin{equation}\label{eq:epsilon_parameter}
\epsilon^2:= \frac{4\left[E - 2\mu^2\left(1 - \frac{4\sigma}{3} + \frac{\sigma^4}{3} \right)\right]}{1+\sigma^2},
\end{equation}
Eq.\eqref{eq:T-independent_Wheeler-DeWitt-Schrodinger} can rewritten as
\begin{equation}\label{eq:Schrodinger}
	\frac{d^2}{da^2}\psi(a)+\frac{\lambda}{a}\frac{d}{da}\psi(a) - V(a)\psi(a)=0, 
\end{equation}
where
\begin{equation}\label{eq:complete_potential}
    V(a):= 4\kappa a^2 - \epsilon^2 a^4.
\end{equation}
Taking the curvature $\kappa =0$ in agreement with the current observations \cite{aghanim_planck_2020}, the solution takes the form
\begin{equation}\label{eq:general_Schrodinger_solution}
	\psi(a)=a^{3\alpha}\left[c_1 H_{\alpha}^{(+)}\left(\frac{\epsilon a^3}{3}\right) + c_2 H_{\alpha}^{(-)}\left(\frac{\epsilon a^3}{3}\right)\right],
\end{equation}
where $\alpha:=\frac{1 - \lambda}{6}$ and where $H_\alpha^{(+)} := H_\alpha^{(1)}$ and $H_\alpha^{(-)} := H_\alpha^{(2)}$ are the Hankel functions of first and second kind, respectively. Its important to note that the definition of $\epsilon$ satisfy the constraint $3E < 2\mu^2(3-4\sigma+\sigma^4)$ \cite{darabi_classical_2016}, which ensures a (bounded) oscillatory wavefunction.

The parameter $\lambda$ do not modify the wavefunction in semiclassical regime. Without the presence of $\lambda$, i.e. for the value $\lambda=0$, Eq.\eqref{eq:Schrodinger} resembles a one-dimensional Schrödinger equation for a particle modeled by a $a(t)$ coordinate with zero energy moving in a potential $V(a)=-\epsilon^2 a^4$, and this is equivalent to the asymptotic approximation \cite{cordero_phase_2014} for values of $a\gg 0$. As is usual in quantum cosmology \cite{vilenkin_approaches_1994}, is precise get $\psi(a)$ in the form $\psi(a)=c_1\psi_{+}(a) + c_2\psi_{-}(a)$, where $\psi_{\pm}(a)$ are the WKB solutions for \eqref{eq:Schrodinger}.

\begin{figure}[t]
	\centering
	\includegraphics[width=0.45\textwidth]{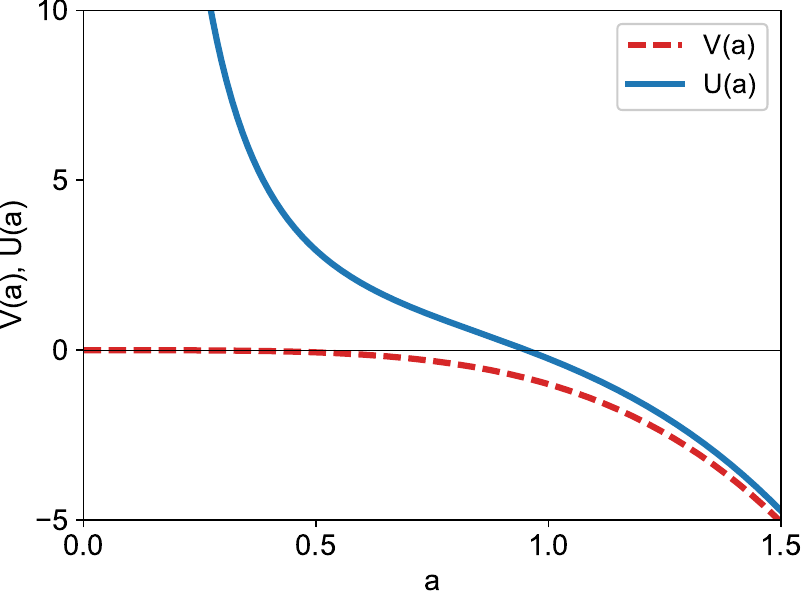}
	\caption{\label{fig:pot} The potential $V(a)= -\epsilon^2 a^4$ and the effective potential $U(a) = -\epsilon^2 a^4-\frac{\lambda}{2a^2}\left(1 - \frac{\lambda}{2}\right)$ used to perform the WKB approximation, for $\epsilon = 1$ and $\lambda = -1$.}
\end{figure}

\subsubsection{\label{sec:WKB_Semiclassical_Approximation}WKB Semiclassical Approximation}

In order to get the WKB description, a suitable potential function is needed. However, the potential $V(a)=- \epsilon^2 a^4$ has no barrier; so, in this case, to get an expression to compare with Eq.\eqref{eq:Schrodinger} is necessary perform the separation $\psi(a) =a^{-\frac{\lambda}{2}}\varphi(a)$. So, it is obtained the equation
\begin{equation}\label{eq:WKB_Schrodinger}
	\frac{d^2}{da^2}\varphi(a) + \left[\epsilon^2 a^4 + \frac{\lambda}{2a^2}\left(1 - \frac{\lambda}{2}\right) \right]\varphi(a)=0,
\end{equation}
with the effective potential
\begin{equation}\label{eq:effective_potential}
    U(a):= -\epsilon^2 a^4 - \frac{\lambda}{2a^2}\left(1 - \frac{\lambda}{2}\right)
\end{equation}
and the turning point at $a_0 = \left[-\frac{\lambda}{2\epsilon^2}\left(1 - \frac{\lambda}{2} \right)\right]^{\frac{1}{6}}$ (as can be seen in Fig. \ref{fig:pot}). In the classical allowed region $a>a_0$ the WKB solutions to Eq.\eqref{eq:WKB_Schrodinger} is given by
\begin{equation}\label{eq:usual_WKB_classical_solution}
	\varphi_{\pm}(a) \approx [p(a)]^{-\frac{1}{2}}\exp\left[\pm i\left( \int_{a_0} ^a d\xi~p(\xi) -\frac{\pi}{4}\right)\right],
\end{equation}
where
\begin{equation}\label{eq:WKB-momentum}
    p(a):=\sqrt{-U(a)}=\sqrt{\epsilon^2 a^4 + \frac{\lambda}{2a^2}\left(1 - \frac{\lambda}{2}\right)} = \frac{\epsilon}{a}\sqrt{a^6 - a_0^6}.
\end{equation}
Taking this into account, it is possible express WKB solutions $\psi_{\pm}(a)$ for $a>a_0$ as
\begin{equation}\label{eq:WKB_classical_solution}
	\psi_{\text{WKB }\pm}(a) \approx a^{-\frac{\lambda}{2}} [p(a)]^{-\frac{1}{2}} e^{\pm i S(a)},
\end{equation}
where
\begin{equation}\label{eq:WKB_action-like_function}
    S(a):= \frac{ap(a)}{3} - \frac{\epsilon a_0^3}{3}\arctan \frac{ap(a)}{\epsilon a_0^3} - \frac{\pi}{4}, 
\end{equation}
and for values of $a\gg a_0$, $p(a) \approx \epsilon a^2$ and
\begin{equation}\label{eq:asymptotic_WKB_classical_solution}
	\psi_{\text{WKB }\pm}(a) \approx \frac{a^{-1-\frac{\lambda}{2}}}{\sqrt{\epsilon}}\exp\left[\pm i\left( \frac{\epsilon a^3}{3} - \frac{\alpha\pi}{2} - \frac{\pi}{4}\right)\right].
\end{equation}

Using the asymptotical expansions for the Hankel functions \cite{zwillinger_table_2015},
\begin{equation}\label{eq:Bessel_asymptotical_expansion}
	H^{(\pm)}_\alpha (z)\approx \sqrt{\frac{2}{\pi z}}e^{\pm i \left( z - \frac{\alpha \pi}{2} - \frac{\pi}{4} \right)},
\end{equation}
it is obtained by comparation between Eq.\eqref{eq:asymptotic_WKB_classical_solution} and Eq.\eqref{eq:general_Schrodinger_solution} that the exact solution in terms of expanding and contracting wavefunctions is
\begin{align}\label{eq:expanding-contracting_wavefunctions}
\psi_{\pm}(a) &= \sqrt{\frac{\pi}{6}} a^{3\alpha} H^{(\pm)}_{\alpha}\left(\frac{\epsilon a^3}{3}\right)\nonumber\\
    & = \sqrt{\frac{\pi}{6}} e^{\pm i\frac{\pi}{2} \left( 1 - \alpha \right) } a^{3\alpha} \frac{e^{\mp i\frac{\alpha\pi}{2}} J_{\alpha}\left( \frac{\epsilon a^3}{3} \right) - e^{\pm i\frac{\alpha\pi}{2}} J_{-\alpha}\left( \frac{\epsilon a^3}{3} \right) }{\sin \alpha\pi}, 
\end{align}
where $J_\alpha$ is the first kind Bessel function. For the characteristic value $\alpha = \frac{1}{3}$, it follows
\begin{equation}\label{eq:lambda_minus_one_expanding-contracting_wavefunctions}
	\psi_{\pm}(a) = \sqrt{\frac{\pi}{2}}\left( \frac{2}{\epsilon} \right)^{\frac{1}{3}} \left[ \AiryAi\left(-\left( \frac{\epsilon}{2} \right)^{\frac{2}{3}} a^2\right) \mp i \AiryBi\left(-\left( \frac{\epsilon}{2} \right)^{\frac{2}{3}} a^2\right)\right],
\end{equation}
where $\AiryAi$ and $\AiryBi$ are the Airy function of first and second kind, respectively.

\begin{figure}[t]
	\centering
	\includegraphics[width=0.45\textwidth]{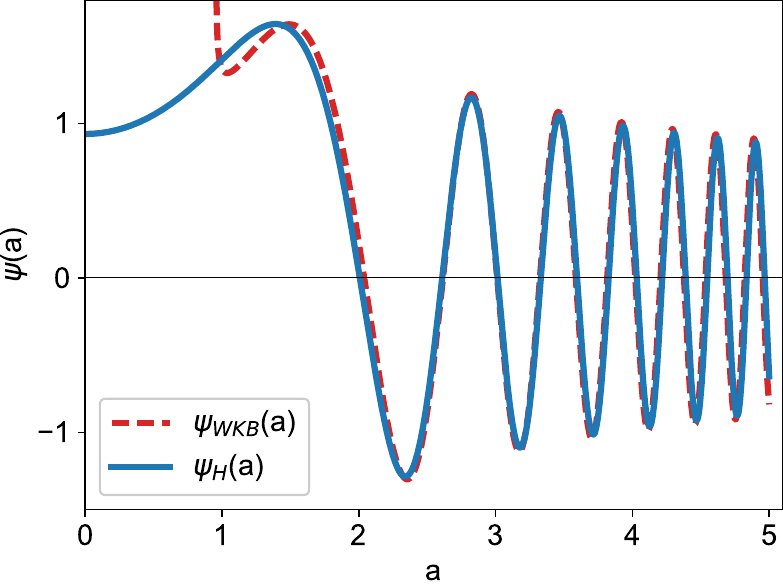}
	\caption{\label{fig:psi} The exact $\psi_H(a)$ and WKB $\psi_{\text{WKB}}(a)$ wavefunctions with Hartle-Hawking boundary conditions (for $\alpha = 1/3, \hslash = 1, \epsilon = 1$).}
\end{figure}

\subsubsection{\label{sec:Boundary_Conditions}Boundary Conditions}

The wavefunction \eqref{eq:expanding-contracting_wavefunctions} need to satisfy specific boundary conditions prescribed by an particular physical environment. In quantum cosmology, there is a boundary condition setting problem \cite{vilenkin_cosmic_2001}, motivated by the different evolution paradigms in cosmology. For this reason, this paper studies three of the most employed boundary conditions: Hartle-Hawking \cite{hartle_wave_1983}, Vilenkin \cite{vilenkin_quantum_1984,vilenkin_approaches_1994} and Linde\cite{linde_quantum_1984} proposals. With this, the aim is analyze some quantum properties of this conditions and their relation with the classical behavior.

\begin{figure}[t]
	\centering
	\includegraphics[width=0.5\textwidth]{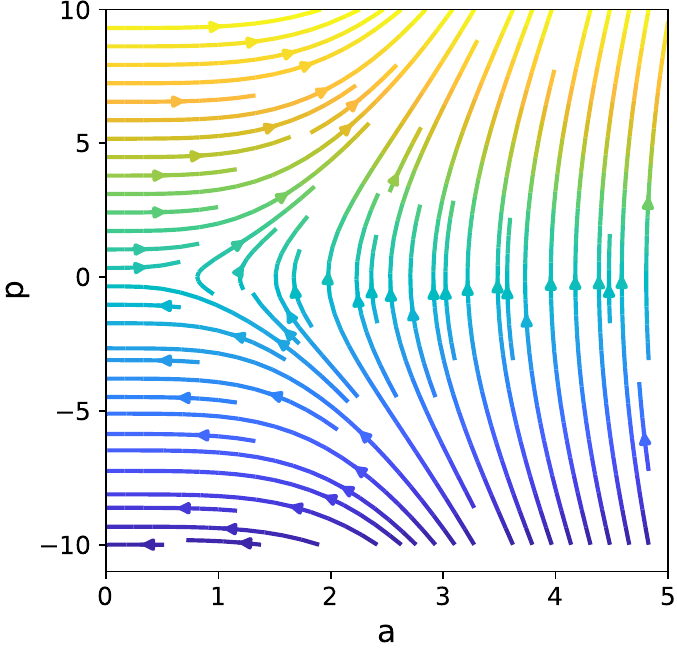}
	\caption{\label{fig:phase} Integral curves plot for the stationary vector field $(\dot a, \dot p) = (2p, 4\epsilon^2a^3)$ in phase space associated to classical Hamiltonian constraint \eqref{eq:sigma-Hamiltonian_constraint}, for $\epsilon = 1$. The integral lines coincides with the loci of extrema of Hartle-Hawking Wigner function.}
\end{figure}

This three boundary conditions in quantum cosmology are

Hartle-Hawking:
\begin{equation}\label{eq:Hartle-Hawking_boundary_coditions}
	\psi_H (a)=\begin{cases}
							\psi_{-}(a) & a<\tau\\
							\psi_{+}(a)+\psi_{-}(a) & a>\tau,
						\end{cases}
\end{equation}

Vilenkin:
\begin{equation}\label{eq:Vilenkin_boundary_conditions}
	\psi_V (a)=\begin{cases}
							\psi_{+}(a) -\frac{i}{2}\psi_{-}(a) & a<\tau\\
							\psi_{-}(a) & a>\tau,
						\end{cases}
\end{equation}

Linde:
\begin{equation}\label{eq:Linde_boundary_conditions}
	\psi_L (a)=\begin{cases}
							\psi_{+}(a) & a<\tau\\
							\psi_{+}(a) - \psi_{-}(a) & a>\tau,
						\end{cases}
\end{equation}
where $\tau$ is the turning point. As discussed above, the semiclassical approximation is not fullfilled, but $\tau$ can be setted as $\tau=0$, so Eqs. \eqref{eq:Hartle-Hawking_boundary_coditions}--\eqref{eq:Linde_boundary_conditions} can be written for $a>0$ as
\begin{equation}\label{eq:Hartle-Hawking_wavefunction}
	\psi_H (a)=\psi_{+}(a)+\psi_{-}(a),
\end{equation}
\begin{equation}\label{eq:Vilenkin_wavefunction}
	\psi_V (a)=\psi_{-}(a),
\end{equation}
and
\begin{equation}\label{eq:Linde_wavefunction}
	\psi_L (a)=\psi_{+}(a) - \psi_{-}(a).
\end{equation}

\begin{figure}[t]
	\centering
	\includegraphics[width=0.5\textwidth]{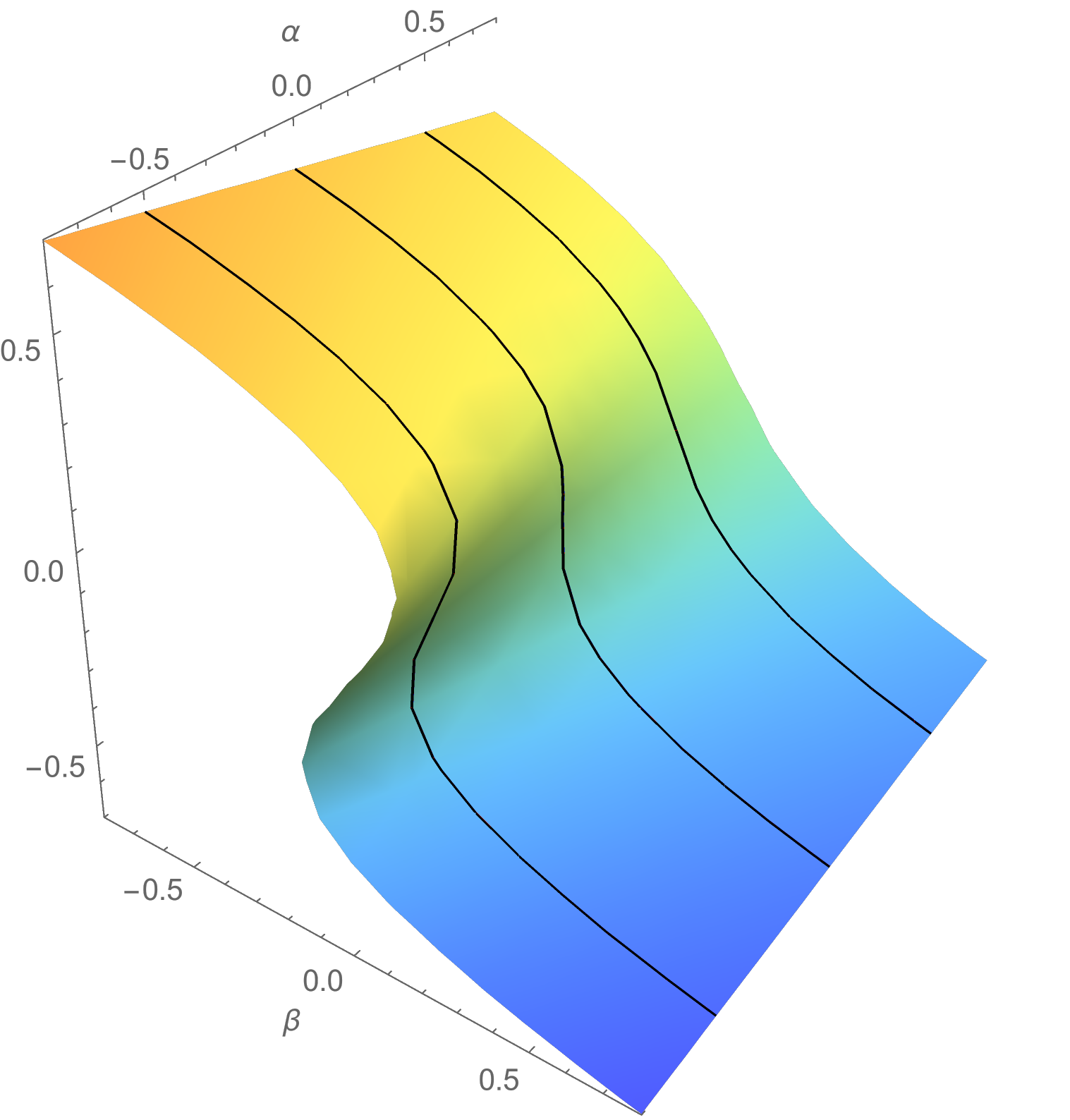}
	\caption{\label{fig:cusp} Cusp catastrophe plot for the potential $V(a) = -\epsilon^2 a^4 + 4\kappa a^2 + \lambda a$, with control term $V_c = \lambda a$ and parameter $\alpha = 4\kappa/\epsilon^2$ and $\beta = \lambda/\epsilon^2$. Trichotomy of curvature $\kappa$ is directly proportional to the stability parameter $\alpha$.}
\end{figure}

In Fig. \ref{fig:psi} the solutions to Wheeler-DeWitt-Schrödinger equation for the Hartle–Hawking boundary condition by means of WKB approximation and Hankel functions are showed. For very large values of $a\gg a_0$ the two solutions are similar. From the WKB wavefunction \eqref{eq:WKB_classical_solution} it follows that
\begin{equation}\label{eq:superclassical_limit_momentum}
	\hat{p}_a\psi_{\pm}(a)\approx \mp p_a\psi_{\pm}(a),
\end{equation}
and taking Eqs. \eqref{eq:Hamilton-a} and \eqref{eq:Hamilton-b} in account, the Eq.\eqref{eq:superclassical_limit_momentum} gives the interpretation for $\psi_{\pm}(a)$ that negative values of $p_a$ correspond to an expanding universe.

Therefore, the Vilenkin wavefunction represents tunneling and includes only a outgoing term corresponding to a expanding universe while Hartle-Hawking and Linde wavefunctions includes expanding and contracting universes with equal weight. In fact, the explicit form of this wavefunctions in WKB approximation are given by
\begin{equation}\label{eq:Hartle-Hawking_WKB_wavefunction}
	\psi_H(a)\approx \frac{2a^{3\alpha}}{\sqrt{\epsilon}\left( a^6-a_0^6 \right)^{\frac{1}{4}}}\cos S(a)=\frac{a^{3\alpha}}{\sqrt{\epsilon}\left( a^6-a_0^6 \right)^{\frac{1}{4}}}(e^{iS(a)}+e^{-iS(a)})
\end{equation}
\begin{equation}\label{eq:Vilenkin_WKB_wavefunction}
	\psi_V(a)\approx \frac{a^{3\alpha}}{\sqrt{\epsilon}\left( a^6-a_0^6 \right)^{\frac{1}{4}}} e^{-iS(a)}
\end{equation}
and
\begin{equation}\label{eq:Linde_WKB_wavefunction}
	\psi_L(a)\approx \frac{2ia^{3\alpha}}{\sqrt{\epsilon}\left( a^6-a_0^6 \right)^{\frac{1}{4}}}\sin S(a)=\frac{a^{3\alpha}}{\sqrt{\epsilon}\left( a^6-a_0^6 \right)^{\frac{1}{4}}}(e^{iS(a)}-e^{-iS(a)}),
\end{equation}

\begin{figure}[t]
	\centering
	\includegraphics[width=0.5\textwidth]{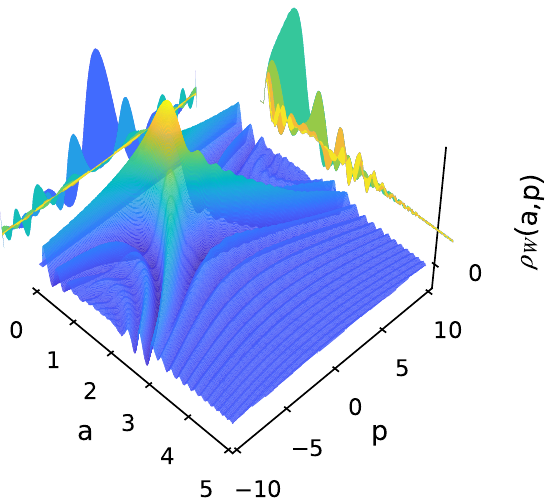}
	\caption{\label{fig:wpH} Hartle-Hawking Wigner function (for $\alpha = 1/3, \hslash=1$, $\epsilon = 1$). The figure shows many oscillations due to the interference between wavefunctions of expanding and contracting universe.}
\end{figure}

where $S(a)$ is given by Eq. \eqref{eq:WKB_action-like_function}. On the other hand, its exact forms are given by
\begin{equation}\label{eq:exact_Hartle-Hawking_wavefunction}
	\psi_H(a) = \sqrt{\frac{\pi}{6}}a^{3\alpha} \left[H^{(+)}_{\alpha}\left( \frac{\epsilon a^{3}}{3} \right) + H^{(-)}_{\alpha}\left( \frac{\epsilon a^{3}}{3} \right)\right],
\end{equation}
\begin{equation}\label{eq:exact_Vilenkin_wavefunction}
	\psi_V (a) = \sqrt{\frac{\pi}{6}}a^{3\alpha} H^{(-)}_{\alpha}\left( \frac{\epsilon a^{3}}{3} \right),
\end{equation}
and
\begin{equation}\label{eq:exact_Linde_wavefunction}
	\psi_L(a) = \sqrt{\frac{\pi}{6}}a^{3\alpha} \left[H^{(+)}_{\alpha}\left( \frac{\epsilon a^{3}}{3} \right) - H^{(-)}_{\alpha}\left( \frac{\epsilon a^{3}}{3} \right)\right].
\end{equation}
As before, for $\alpha = \frac{1}{3}$,
\begin{equation}\label{eq:lambda_minus_one_exact_Hartle-Hawking_wavefunction}
	\psi_H(a) = \sqrt{2\pi}\left( \frac{2}{\epsilon} \right)^{\frac{1}{3}} \AiryAi\left( -\left( \frac{\epsilon}{2} \right)^{\frac{2}{3}} a^2\right),
\end{equation}
\begin{equation}\label{eq:lambda_minus_one_exact_Vilenkin_wavefunction}
	\psi_V (a) = \sqrt{\frac{\pi}{2}}\left( \frac{2}{\epsilon} \right)^{\frac{1}{3}}\left[\AiryAi\left( -\left( \frac{\epsilon}{2} \right)^{\frac{2}{3}} a^2\right) + i\AiryBi\left( -\left( \frac{\epsilon}{2} \right)^{\frac{2}{3}} a^2\right)\right],
\end{equation}
and
\begin{equation}\label{eq:lambda_minus_one_exact_Linde_wavefunction}
	\psi_L(a) = -i \sqrt{2\pi}\left( \frac{2}{\epsilon} \right)^{\frac{1}{3}}\AiryBi\left( -\left( \frac{\epsilon}{2} \right)^{\frac{2}{3}} a^2\right).
\end{equation}
Its important to note that there is an open discussion about which could be the correct boundary condition in quantum cosmology. An exploration of this problem in the context of deformation quantization, a formalism beyond canonical quantization, will be developed in next section.

\begin{figure}[t]
	\centering
	\includegraphics[width=0.45\textwidth]{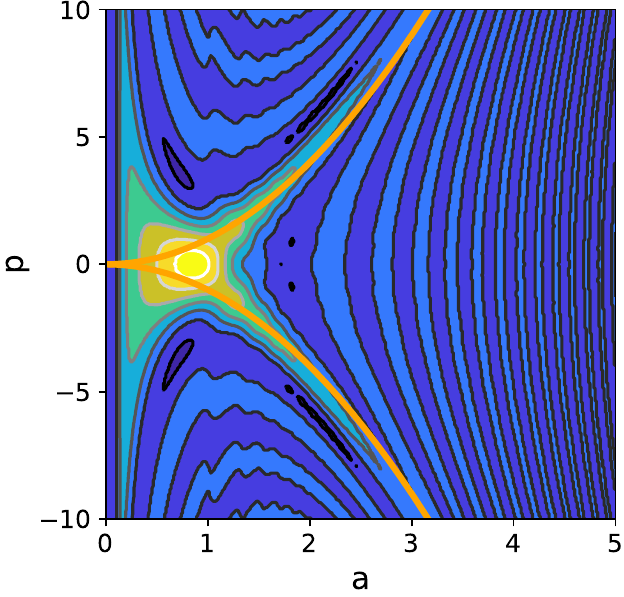}
	\caption{\label{fig:dpH} Hartle-Hawking Wigner function contour-density projection plot (for $\alpha = 1/3, \hslash = 1$, $\epsilon = 1$). It is observed that the classical trajectory coincides with the highest peaks of its corresponding Wigner function.}
\end{figure}

\subsection{\label{sec:Deformation_Quantization_Minisuperspace}Deformation Quantization in the Minisuperspace}
The next step in deformation quantization is apply the non-commutative Moyal $\star$-product in the construction of Wheeler-DeWitt-Moyal equation \eqref{eq:Wheeler-DeWitt-Moyal} in order to get the Wigner functional in a minimal bigravity model. The exposition in previous section put the focus on functional integrals defined in the whole superspace of three-metrics; however, it is unnecessary take this approach because the initial restriction to work at the minisuperspace.

\subsubsection{\label{sec:Exact_Solution_Hartle-Hawking_Case}Exact Solution: Hartle-Hawking Case}
With the restriction of work in minisuperspace it is possible to search the Wigner function and rewrite the Hamiltonian constraint \eqref{eq:massive_Hamiltonian_constraint} in terms of the Moyal $\star$-product and the Wigner function as
\begin{equation}\label{eq:Moyal_Hamiltonian_constraint}
	\left[-(1+\sigma^2)\frac{p_a^2}{4a} - (1+\sigma^2)\kappa a+ \frac{E}{a^{3\omega}} - 2\mu^2a^3\left(1-\frac{4}{3}\sigma+\frac{\sigma^4}{3}\right)\right]\star\rho_W(a,p_a)=0,
\end{equation}
where the corresponding Moyal $\star$-product is given by
\begin{equation} \label{eq:Moyal_star_product_functions}
	\star=\exp\left\{\frac{i\hslash}{2}\overleftrightarrow{\mathcal P}\right\}=\exp\left\{\frac{i\hslash}{2}\left(\frac{\overleftarrow{\partial}}{\partial a}\frac{\overrightarrow{\partial}}{\partial p_a} - \frac{\overleftarrow{\partial}}{\partial p_a}\frac{\overrightarrow{\partial}}{\partial a}\right)\right\}
\end{equation}
or recalling the assumptions made it to get Eq. \eqref{eq:Schrodinger}, aside the change of variables $p:=p_a$, $q:=\epsilon^2 a^4$,
\begin{equation}\label{eq:Wheeler-DeWitt-Moyal_functions}
	\mathcal H_0 \star\rho_W(q,p)=\left(p^2-q\right)\star\rho_W(q,p).
\end{equation}
Using the Bopp shift relation \cite{zachos_quantum_2005},
\begin{equation}\label{eq:Bopp}
f(q,p)\star g(q,p)=f\left(q + \frac{i\hslash}{2}~\overrightarrow{\partial}_p,p - \frac{i\hslash}{2}~\overrightarrow{\partial}_q\right)g(q,p),
\end{equation}
results in
\begin{equation}\label{eq:Liouville-Moyal}
	\left(p^2 - i\hslash p\partial_q - \frac{\hslash^2}{4}\partial_q^2 - q - \frac{i\hslash}{2} \partial_p \right)\rho_W (a,p)=0.
\end{equation}

\begin{figure}[t]
	\centering
	\includegraphics[width=0.5\textwidth]{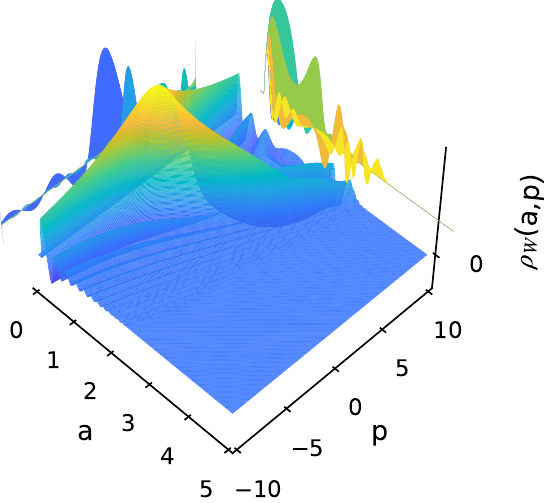}
	\caption{\label{fig:wpV} Vilenkin Wigner function (for $\alpha = 1/3, \hslash=1$, $\epsilon = 1$). The figure shows a clear maximum and less oscillations compared with the Hartle-Hawking Wigner function.}
\end{figure}

If a new variable $z$ is defined as $z:=p^2 - q$, then the imaginary part of Eq.\eqref{eq:Liouville-Moyal} is identically equals to zero and the real part takes the form
\begin{equation}\label{eq:Airy-Moyal}
	\frac{d^2}{dz^2}\rho_W - \frac{4}{\hslash^2} z \rho_W=0,
\end{equation}
whose solution is
\begin{equation}\label{eq:Wigner_function}
	\rho_W(a,p)=c_1 \AiryAi\left( \left(\frac{2}{\hslash}\right)^{\frac{2}{3}}(p^2 - \epsilon^2 a^4) \right).
\end{equation}
Eq. \eqref{eq:Airy-Moyal} indeed admits another solution corresponding to $\AiryBi\left( \left(\frac{2}{\hslash}\right)^{\frac{2}{3}}(p^2 - \epsilon^2 a^4) \right)$, but in this case, the Airy function of second kind $\AiryBi(x)$ is divergent for $x>0$ and this part cannot be included as a suitable Wigner function. With the purpose of obtaining the next result, it is necessary the assumption of the existence of the Wigner transform. As is pointed in Refs. \cite{cordero_deformation_2011,akhundova_wigner_1992}, there is exist a problem with the range of integration over all real line for the variable $q$ (and then the scale factor $a$). So, in order to avoid this and get a solution to \eqref{eq:Wheeler-DeWitt-Moyal}, it is convenient use the following integral representation of Wigner function:
\begin{equation}\label{eq:integral_Wigner_function}
	\rho_W (a,p)=\frac{1}{\pi\hslash}\int_{-\infty}^{\infty}~d\xi \exp\left\{-2i\frac{\xi p}{\hslash}\right\}\psi^*(a-\xi) \psi(a+\xi).
\end{equation}

Using the convolution theorem and the Fourier transform for the Airy function is possible get the Wigner function for the Hartle-Hawking wavefunction
\begin{equation}\label{eq:Wigner-Hartle-Hawking_function_exact_integral}
	\rho_W (a,p) = \frac{3}{\sqrt{2\pi} \epsilon^{\frac{2}{3}}}\AiryAi\left( \left(\frac{2}{\hslash}\right)^{\frac{2}{3}}(p^2 - \epsilon^2 a^4) \right).
\end{equation}

\begin{figure}[t]
	\centering
	\includegraphics[width=0.45\textwidth]{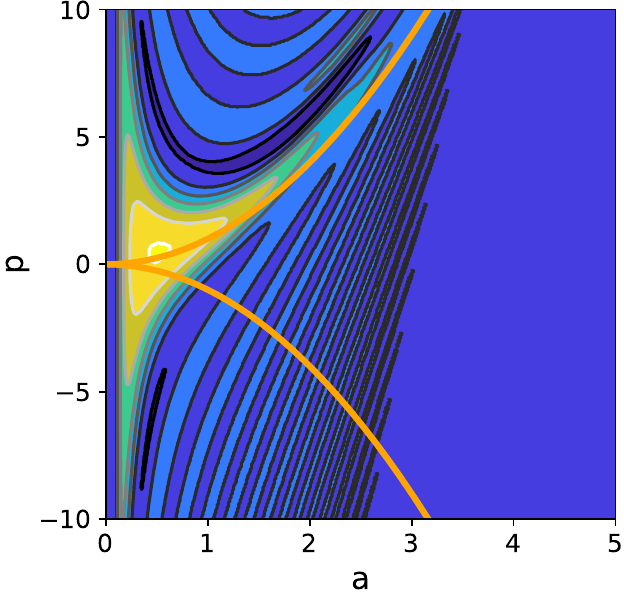}
	\caption{\label{fig:dpV} Vilenkin Wigner function contour-density projection plot (for $\alpha = 1/3, \hslash =1$, $\epsilon = 1$). It is observed that the classical trajectory is at some points on the highest peaks of its corresponding Wigner function and has only one branch.}
\end{figure}

Eq.\eqref{eq:Wigner-Hartle-Hawking_function_exact_integral} is the principal object in deformation quantization: essentially, provides all the physical information of the system, in this case, the dynamics of a universe in a minimal bigravity in minisuperspace phase space. To describe the coincident classical trajectories and compare with classical behavior \cite{steuernagel_wigner_2013,bernardini_quantum_2018}, is necessary to consider the Hamilton equations for $a$ and $p$ associated to classical Hamiltonian constraint \eqref{eq:sigma-Hamiltonian_constraint} (for $p_T\mapsto E$, $\kappa = 0$ and $\omega = -1$):
\begin{align}
	\dot a &= \{a,\mathcal H_0\}_{\text{PB}} = 2p \label{eq:classical_Hamiltonian_a_PB}\\
	\dot p &= \{p,\mathcal H_0\}_{\text{PB}} = 4\epsilon^2 a^3.\label{eq:classical_Hamiltonian_p_PB}
\end{align}
Eqs. \eqref{eq:classical_Hamiltonian_a_PB} and \eqref{eq:classical_Hamiltonian_p_PB} defines the stationary vector field in phase space $(\dot a, \dot p) = (2p, 4\epsilon^2a^3)$, which is showed in Fig. \ref{fig:phase}. The locus \cite{rashki_quantum_2017,calzetta_wigner_1989} of extrema of Wigner function \eqref{eq:Wigner-Hartle-Hawking_function_exact_integral} is the deformed version of the classical Hamiltonian given by
\begin{equation}\label{eq:maxima_locus_Hartle-Hawking}
	p^2 - \epsilon^2 a^4 + \left( \frac{\hslash}{2} \right)^{\frac{2}{3}}a'_n = 0,
\end{equation}
where $a'_n$ is the $n$-th zero of the derivative of Airy function, $d/dx\AiryAi(-x)|_{x = a'_n} = 0$ (see Table \ref{tab:Airy}). Eq. \eqref{eq:maxima_locus_Hartle-Hawking} presents the most probable cosmological solutions.

\begin{table}[b]
\caption{Zeroes $a_n$of $\AiryAi(-x)$  and zeroes $a'_n$ of $\frac{d}{dx}\AiryAi(-x)$ for $n = 1, 2, 3, 4$}
\label{tab:Airy}
\centering
	\begin{tabular}{ccc}\\\hline\hline
		$n$ & $a_n$ & $a'_n$\\\hline
		1 & 2.33811\dots & 1.01879\dots \\
		2 & 4.00795\dots & 3.24819\dots \\
		3 & 5.55056\dots & 4.82009\dots\\
		4 & 6.78671\dots & 6.16331\dots
	\end{tabular}
\end{table}

Before continue with Vilenkin and Linde cases, some remarks are necessary about Hartle-Hawking Wigner function. As is pointed in \cite{berry_semi-classical_1977}, Wigner functions may exhibit a cusp catastrophe behavior in nonclassical regions of phase space. Catastrophes of $\rho_W$ occurs at points $(a,p)$ where the stationary phase approximation diverges, and resembles the caustics of families of trajectories in configuration or momentum space, although, due to Liouville theorem, there are no caustics in phase space. If the complete potential \eqref{eq:complete_potential} is extended with a control term $V_c = \lambda a$, it obtains the prototypical cusp catastrophe potential
\begin{equation}\label{eq:cusp_potential}
	V(a) = -\epsilon^2 a^4 + 4\kappa a^2 + \lambda a
\end{equation}

\begin{figure}[t]
	\centering
	\includegraphics[width=0.5\textwidth]{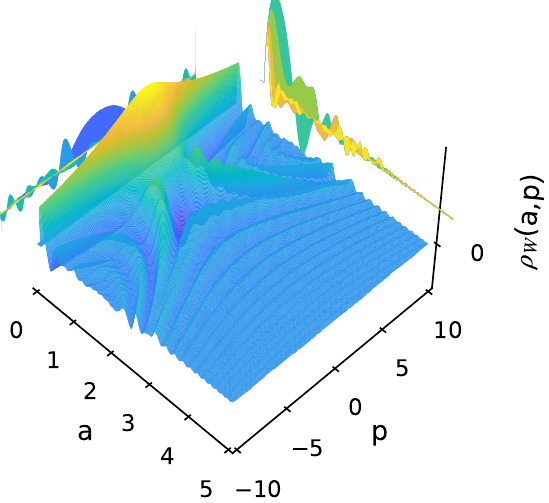}
	\caption{\label{fig:wpL} Linde Wigner function (for $\alpha = 1/3, \hslash = 1$, $\epsilon = 1$). The figure shows many oscillations due to the interference between wavefunctions of expanding and contracting universe.}
\end{figure}

In Fig. \ref{fig:cusp} is showed the cusp catastrophe plot, a surface where $a$ satisfying $dV(a)/da = 0$ for parameters $\alpha = 4\kappa/\epsilon^2$ and $\beta = \lambda/\epsilon^2$, drawn for parameter $\beta$ continuously varied and for several values of parameter $\alpha$. Outside the cusp locus of bifurcations, for each point $(\alpha,\beta)$ in parameter space there is only one extreme value of $a$. Inside the cusp, there are two different values of $a$ giving local minima of $V(a)$ for each $(\alpha,\beta)$, separated by a value of $a$ giving a local maximum. From these parameters, it is possible get a qualitative bound for $\kappa$ and $\epsilon$ in terms of stability criteria. A qualitative dimensional comparation between \eqref{eq:cusp_potential} and \eqref{eq:maxima_locus_Hartle-Hawking} suggest that $V_c$ can be related with $\left( \frac{\hslash}{2} \right)^{\frac{2}{3}} a_n$ and then $\lambda = \left( \frac{\hslash}{2} \right)^{\frac{2}{3}}$. On the other hand, the trichotomy of parameter $\alpha = 4\kappa/\epsilon^2$ ($\alpha < 0$ for two unstable solutions, $\alpha = 0$ for bifurcation point and $\alpha > 0$ for one stable solution) shows that in this cosmological model, definite-positive values for curvature $\kappa$ implies stable solutions, giving another argument for the election of flat geometries in a minimal bigravity theory.

\begin{figure}[t]
	\centering
	\includegraphics[width=0.45\textwidth]{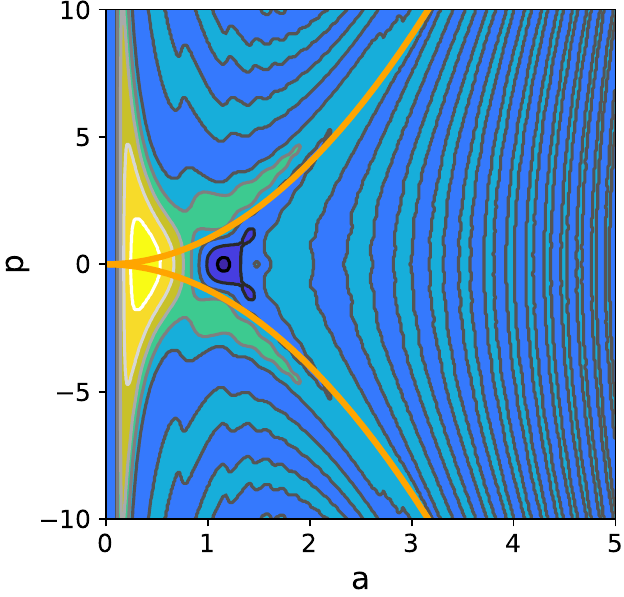}
	\caption{\label{fig:dpL} Linde Wigner function contour-density projection plot (for $\alpha = 1/3, \hslash = 1$, $\epsilon = 1$). It is observed that the classical trajectory coincides with the highest peaks of its corresponding Wigner function, similarly as in Hartle-Hawking case.}
\end{figure}

\subsubsection{\label{sec:Numerical_Analysis}Numerical Analysis of Wigner Function}
Eq. \eqref{eq:Wigner-Hartle-Hawking_function_exact_integral} is obtained integrating out from minus to plus infinity in variable $a$, even though the wavefunction only is defined for positive values of the scale factor. Because the exponential decay of Airy function $\AiryAi(x)$ for $x>0$, the wavefunction and its associated Wigner function are very similar. As is mentioned in \cite{akhundova_wigner_1992}, the restriction to positive values for the calculation of an analytical expression from Eq. \eqref{eq:integral_Wigner_function} is very difficult even using the WKB approximation, so it is necessary perform a numerical integration. Through an implementation in Python language the Wigner functions for Hartle-Hawking, Vilenkin, and Linde wavefunctions are calculated and the results depicted in Figs. \ref{fig:wpH}, \ref{fig:wpV}, and \ref{fig:wpL}. Their respective density-contour plots are showed in Figs. \ref{fig:dpH}, \ref{fig:dpV}, and \ref{fig:dpL}, respectively.

After obtaining the Wigner functions for the different boundary conditions, some remarks are necessary. All this plots can be described by the WKB approximation for the wavefunctions, as it is pointed out in Refs. \cite{habib_wigner_1990,cordero_deformation_2011,cordero_phase_2019}.

For the Hartle-Hawking Wigner function an oscillatory symmetric form around zero momentum is observed (see Fig. \ref{fig:wpH}), where its largest peak corresponds to the classical trajectory given by
\begin{equation}\label{eq:classical_trajectory}
	p^2 =\epsilon^2 a^4,
\end{equation}
which are represented by thick lines in the density plots (see Figs. \ref{fig:dpH}, \ref{fig:dpV}, and \ref{fig:dpL}) for the characteristic value $\epsilon = 1$. It is observed that the extrema (maxima and minima) of Wigner function decrease with the distance to the classical trajectory. Besides the Wigner function has negative values which are interpreted by some authors \cite{lutkenhaus_nonclassical_1995,kenfack_negativity_2004} as nonclassical effects (similar characteristics are presented in the other cases).

The Vilenkin Wigner function (see Fig. \ref{fig:wpV}), is constructed with the wavefunction corresponding to an expanding universe. In this case there are not interference terms, as the presented in Hartle-Hawking case. This behavior corresponds to the prototypical tunneling process. For this reason the highest peaks of the Wigner function are near to the classical trajectory associated with negative values of the momenta (expanding universe; see Fig. \ref{fig:dpV}).

The Linde Wigner function (see Fig. \ref{fig:wpL}) has a similar behavior as the Hartle-Hawking case, but the classical trajectory is not on the highest peaks (in fact, seems that is on the minima of Wigner function), as it is showed in Fig. \ref{fig:dpL}. More fluctuations than in Hartle-Hawking case are observed, mainly on the left of the region bounded by the classical trajectory, meanwhile outside this region the peaks of the oscillations damped drastically their amplitudes.

For Hartle-Hawking and Linde cases the Wigner function have interference terms corresponding to a contracting and expanding universe; the difference between them is an opposite sign in the interference term of the wavefunctions \eqref{eq:Hartle-Hawking_boundary_coditions} and \eqref{eq:Linde_boundary_conditions}.

From point of view of quantum cosmology, this results suggest that the classical limit is difficult to achieve since there exist oscillations of the Wigner function for the three cases considered. Until the classical trajectory is on the highest peaks in Hartle-Hawking Wigner function, the decoherence of the Vilenkin function could be easier to obtain because it has fewer oscillations to the other cases due the interference terms not being present since there is only an expanding universe around $p=0$. Besides, with the presence of a tunneling scenario, it is possible avoid the classical initial singularity by means the probability for creation from nothing at a small nonvanish scale factor. By extremizing the marginal
\begin{equation}\label{eq:Wigner_marginal}
	|\psi(a)|^2=\int_{-\infty}^{\infty}dp~\rho_W(a,p),
\end{equation}
it is found that the size of initial scale factor corresponding to the maximum contribution of $|\psi(a)|^2$ is determinated by the frequency term $\epsilon$ and, as it can seen from Eq. \eqref{eq:epsilon_parameter}, by the mass of graviton $m$. Explicitly,
\begin{equation}\label{eq:mass_graviton}
	m = \frac{G m_P}{\sqrt{2}}\sqrt{\frac{E - (1 + \sigma^2)a_1^{-3}}{1-\frac{4\sigma}{3} + \frac{\sigma^4}{3}}},
\end{equation}
where $G$ is the gravitational constant, $m_P$ is the Plank mass and $a_1$ is the first zero of Airy function $\AiryAi(-x)$ (see Table \ref{tab:Airy}). In the context of minimal bigravity, the mass of graviton besides play a role of cosmological constant \cite{darabi_classical_2016} (taking a constant $\sigma = \frac{b}{a} \neq 1$ in account). So, in concordance with \cite{habib_wigner_1990}, the deviation of classical behavior of Wigner functions is in function of the cosmological constant and therefore, in function of $m$. In addition, Sasaki \textit{et al.} \cite{sasaki_hartlehawking_2013} have shown that at early universe for Hartle-Hawking boundary conditions, dRGT massive gravity give a dynamical character to the mass of graviton. They also shown that massive gravity sector improve the probability of a large number of e-foldings for a sufficiently large graviton mass comparable with the Hubble parameter during inflationary stage of order $m\approx 10^{12}$ GeV. 

\section{\label{sec:Conclusions}Conclusions}
The phase space quantum description of a physical system is a construction where the coordinates and momenta coexists simultaneously and in where some physical properties can be appreciated in a more complete form with respect to a particular representation. The key piece in this approach is the Wigner function which contains all quantum information in the phase space. The Wigner quasi-probabilistic distribution and their properties are useful in the description of semiclassical effects and to obtain the classical limit.

In this paper the quantum phase space description was employed to analyze the physical behavior of a nonlinear massive minimal bigravity model. In sake of simplification, Friedmann-Lemaître-Robertson-Walker-like metrics with common curvature $\kappa$ were supposed. Due the presence of interaction between two scale factors in the massive bigravity action and by means of momentum constraint, it was possible the reduction of the two-variable minisuperspace into a one-variable minisuperspace including one scale factor, $a$. With the particular selection of perfect fluid $\omega=-1$ and flat curvature $\kappa=0$, corresponding to an early universe, the Wheeler-DeWitt-Schrödinger and Wheeler-DeWitt-Moyal equations were obtained. For the Wheeler-DeWitt-Schrödinger an exact and a WKB solution were found, meanwhile by numerical analysis the Wigner function associated to a three distinct types of boundary conditions, Hartle-Hawking, Vilenkin and Linde were calculated, and also an exact analytical Wigner function was found for the Hartle-Hawking boundary case.

The behavior of the Hartle-Hawking Wigner function has an oscillatory symmetric form around zero momentum and presents high frequency interference terms between expanding and contracting universe, with the highest peak corresponding with the classical trajectory.

The Linde Wigner function shows a very similar behavior with the Hartle-Hawking case, where there are also expanding and contracting terms of the wavefunction. As it can seen, there are fewer oscillations outside the region of phase space bounded by the classical trajectory, interchanging maxima with minima. This behavior is due to the sign difference in the interference term, and it has as a consequence that the classical trajectory is not on the highest peak of the Wigner function.

For the Vilenkin tunneling boundary condition its associated Wigner function shows oscillations that rapidly damp since there is only one term of expanding universe, that can be explained because there are non-interference terms. Besides, the classical trajectory is closer to the maxima of Wigner function and the process of decoherence seems to be feasible to obtain because the damping of interference terms.

These results indicates that in a minimal bigravity model of universe the classical limit is difficult to obtain for the three boundary conditions as a consequence of the oscillations in Wigner function, despite the Vilenkin Wigner function could be simpler to obtain because there is only an expanding universe. In addition, a relation between this oscillations with frequency $\epsilon$ and the mass of graviton $m$ through the cosmological constant interpretation indicates the role of the mass in quantum cosmological bigravity models. Also, a relation between curvature $\kappa$ and stability of Wigner function through catastrophe cusp plot found, giving an argument to discriminate possible geometries. 

The phase space quantum analysis developed in this work drive the study to other minimal bigravity models where the curvature are not flat or where the topology of space is nontrivial \cite{levin_topology_2002,luminet_status_2016}. For example, it can be applied to analyze the properties in changes of topology \cite{zeldovich_quantum_1984,de_lorenci_topology_1997,vilenkin_cosmic_2001} by means of the Wigner function, where the case of the different boundary conditions can be worked with more detail.

Besides, deformation quantization formalism is up to now the most general method to quantize arbitrary systems \cite{bayen_deformation_1978}, and could guide to obtaining analytical expressions for the Wigner function of a wide class of cosmological objects. The previous approach could allow new results and some would be analyzed in the classical limit in a more systematic way. Both the nontrivial topology spaces cosmology quantization an the issues corresponding to the classical limit are the subject to current research. In future works eventually it will discussed the application of deformation quantization formalism to curved phase spaces and other spaces without trivial topology, and an research of this work under the Wigner flow analysis \cite{steuernagel_wigner_2013,bernardini_quantum_2018,rashki_quantum_2017} is planned as a natural extension.

\section*{Acknowledgements}
The author want to thank the referee for the comments and suggestions which allowed to improve this work. The author is indebted to Rubén Cordero and Marta Costa for all his help and fruitful discussions.

\end{document}